\documentclass[aip,jcp,amsmath,amssymb]{revtex4-1}

\usepackage{graphicx}
\usepackage{amsmath}
\usepackage{amssymb}
\usepackage{ifthen}
\usepackage{booktabs}

\usepackage{graphicx}% Include figure files
\usepackage{dcolumn}% Align table columns on decimal point
\usepackage{bm}% bold math
\usepackage{longtable}
\usepackage{gensymb}
\usepackage[normalem]{ulem}
\usepackage{color}

\begin{document}

%\preprint{}

\title{Orientational ordering and phase behaviour of a binary mixture of
hard spheres and hard spherocylinders} %Title of paper

%repeat the \author .. \affiliation  etc. as needed


\author{Liang Wu}
\affiliation{Department of Chemical Engineering,
Imperial College London, South Kensington Campus, London, SW7 2AZ, United Kingdom}

\author{Alexandr Malijevsk\'{y}}
\affiliation{Department of Physical Chemistry, ICT Prague, 166 28, Praha 6, 
Czech Republic and Institute of Chemical Process Fundamentals of ASCR, 16502 Praha 6,
Czech Republic}

\author{George Jackson}
\author{Erich A. M\"{u}ller}
\affiliation{Department of Chemical Engineering,
Imperial College London, South Kensington Campus, London, SW7 2AZ, United Kingdom}

\author{Carlos Avenda\~{n}o}\email{Corresponding author; carlos.avendano@manchester.ac.uk}
%\affiliation{Department of Chemical Engineering,
%Imperial College London, South Kensington Campus, London, SW7 2AZ, United Kingdom}
\affiliation{School of Chemical Engineering and Analytical Science,
The University of Manchester, Sackville Street,Manchester M13 9PL,United Kingdom}

%\thanks{}


\date{\today}

\begin{abstract}

We study structure and fluid-phase behaviour of a binary mixture of hard spheres
(HSs) and hard spherocylinders (HSCs) in isotropic and nematic states using the
$NP_nAT$ ensemble Monte Carlo (MC) method in which a normal pressure
tensor component is fixed in a system confined between two hard walls. The
method allows one to  estimate the
location of the isotropic-nematic phase transition and to observe the 
asymmetry in the composition 
between the coexisting  phases, with the expected increase
of the HSC concentration in the nematic phase. This is in stark contrast with
the previously reported MC simulations where a conventional isotropic
$NPT$ ensemble was used. We further compare the simulation results with the
theoretical predictions of two analytic theories that extend the original
Parsons-Lee theory using the one-fluid and the many-fluid approximation
[Malijevsk\'y {\it at al} J. Chem. Phys. \textbf{129}, 144504 (2008)]. In
the one-fluid version of the theory the properties of the mixture are mapped on an effective one-component HS
system while in the many-fluid theory the components of the
mixtures are represented as separate effective HS particles.
The
comparison reveals that both the one- and the many-fluid approaches provide a
reasonably accurate quantitative description of the mixture including the
predictions of the isotropic-nematic phase boundary and degree of orientational
order of the HSC-HS mixtures.
\end{abstract}

\pacs{}% insert suggested PACS numbers in braces on next line

\maketitle %\maketitle must follow title, authors, abstract and \pacs

\section{Introduction}\label{chp5_Sec_Intro}

%experiment%simulation
Advanced materials formed by the self-assembly of non-spherical building blocks
has experienced an unprecedented growth due to recent advances in experimental
techniques to create nano and colloidal particles with
almost any imaginable shape. \cite{kimjacs06,herjpcc07,saccocis11,sacnat13}
Functional materials can be engineered by tailoring
the properties of the individual building blocks. \cite{xiaam01,bonmrs05}
Colloidal particles are particularly attractive as building blocks for the
design of mesoscale materials, which are difficult to fabricate by using chemical synthesis,
as  their interactions can also be modulated by modifying both the surface chemistry
of the particles as well and the properties of the solvent media. \cite{bonmrs05}
It is possible to tune the interactions of either steric-stabilised or
charge-stabilised colloidal particles as nearly as hard body-like by matching
the index of refraction of both particles and solvent.
\cite{yetnat03}
Moreover, the self-assembly of these systems can also be controlled by the aid
of external forces such as magnetic and electric fields, gravity \cite{savpre04}, and even
the use of geometrical confinement.
\cite{pieprl83,schprl96,forjpcm06,lowjpcm09,ramsm09,rillangmuir10,grzacsnano10,wenepj13,avesm13}
We refer to these processes in general
as directed self-assembly. \cite{fursm13}

Anisotropic particles can exhibit many fascinating structures in bulk and
confinement including crystals, plastic crystals, and liquid crystal (LC) phases.
\cite{gloscience07,aganm11,damscience12}
LC phases in rod-like particles, for example, are observed in many natural and anthropogenic systems.
Examples include
suspensions of colloidal particles such as vandium pentoxide ($\rm
{V_{2}O_{5}}$) \cite{zoczac25},
and Gibbsite ($\rm Al(OH)_{3}$) \cite{koonat00,Wensink2009}, carbon nanotubes \cite{Bravo2010Carbon},
and some biological systems such as protein fibers \cite{Mezzenga2010},
tobacco mosaic virus \cite{Parsegian1976N}, \textit{fd}-virus
\cite{Dogic1997PRL,Dogic2000Lang,Grelet2003PRL,Purdy2005,Varga2005,Dogic2006COCIS,grelet2014hard},
polypeptide solutions \cite{Miller1974,Horton1990N}, and DNA \cite{Nakata2007S}.
Over the years, simple but non-trivial hard-core models have been used to study the
formation of LC phases. \cite{Allen1993} These models have played an important role to
understand the behaviour of real systems. In particular, the hard-spherocylinder 
(HSC) has been used as a standard model to describe the LC behaviour of rod-like colloidal particles. The HSC
model consist of a cylinder of length $L$ and diameter $D$ capped at each end by a hemisphere
of diameter $D$, and it is shown in Fig. \ref{fig1}(a). Depending of the aspect ratio of the model, corresponding to the ratio $L/D$,
suspensions of HSCs can exhibit the formation
of isotropic, nematic, smectic, and solid phases. This rich phase behaviour has been confirmed by
extensive computer simulations.
\cite{Stroobants1986PRL,Stroobants1987PRA,Veerman1990PRA,McGrother1996,Bolhuis1997JCP}

Despite our profound knowledge on the phase behaviour of rod-like particles, our
understanding of the phase behaviour of  mixtures of anisotropic colloidal particles is
still limited due to the large parameter space that has to be explored, i.e., different
combinations of concentrations, shapes, and sizes of the components, as well as different
thermodynamic conditions. Experimental studies for mixtures of rod-like and spherical
colloids have been reported.
\cite{koelangmuir99,Vliegen,klupre00,helprl03,yassm10,guujpcm12,yenl13,ahmlan14} From the
modelling perspective, binary mixtures of  HSC particles have been studied using
computer simulations including rod-rod \cite{vanphysicaa98,Purdy2005,Varga2005,oyajcp15}, rod-disc
\cite{Kooij2000PRL,Galindo2003JCP,Cuetos2008PRL} and rod-sphere
\cite{Adams1998Nature,Shoot2002JCP,boljpcm03,Lago2004JMR,schjcp04,cuepre07,schprl07,avecpl09,yenl13,Oyarzun2013JCP} systems.
These mixtures have shown the possibility of forming new structures with properties which
are
difficult to attain in pure component mixtures. Rod-sphere mixtures are of particular
interest as this case corresponds to one of the simplest colloid mixtures models, and the
additional possibility of purely entropic depletion interactions, which can give rise to
rich phase phenomena depending on the relative size ratio between the rods and
the spheres.\cite{Adams1998Nature,Vilegenthart1999JCP,Urakami2003JCP,Chen2004,Oversteegen2004JPC}

%theory
The first statistical-mechanical theory to describe the isotropic-nematic phase transition of liquid crystal models was developed by Onsager. In his seminal work,
Onsager \cite{onspr42,Onsager1949,vrorpp92} derived his simple density functional theory (DFT) for the isotropic-nematic transition by truncating the virial expansion
at the level of second virial coefficient. The equilibrium state can then be determined by functional variation of the free energy with respect to the orientational
distribution function. Although Onsager's description is shown to be exact when the rods become infinitely long (because higher-order virial coefficients become
negligible decaying as $D/L$ \cite{Frenkel1987JPC}), the theory does not accurately describe the phase behaviour of rod-like systems of intermediate values of $L/D$
when higher-order virial contributions are neglected. Several attempts have been made to extend Onsager's theory by including the higher-order interactions. Recent
progress in DFT \cite{Vroege1992,Schmidt2001PRE,Brader2002PRE,Esztermann2006,Hansen-Goos2009,HensenGoos2010JPCM,Lai2010JCP} can provide appropriate approaches to the
predictions of the thermodynamic properties of anisotropic fluids. A new free energy functional for inhomogeneous anisotropic fluids of arbitrary shape have been
proposed within the framework of fundamental-measure theory \cite{Schmidt2001PRE} which is based upon careful analysis of the geometry of the particles.
Alternatively, the Parsons-Lee \cite{Parsons1979,Lee1987,Lee1988} approach provides a simple yet efficient way to incorporate the higher-order virial contributions
which is neglected in Onsager's method. Parsons \cite{Parsons1979} proposed an approximation to decouple the orientational and translational degrees of freedom by
mapping the properties of the rods to those of a reference HS system. Lee \cite{Lee1987,Lee1988} approached the problem in a different way by
introducing a scaling relation between virial coefficients of anisotropic particles and HS reference. Following two separate routes, Parsons and Lee reached the same
expression for the free energy functional which is commonly known as the Parsons-Lee (PL) theory \cite{Vroege1992,McGrother1996,Wensink2009,Gamez2010,Wu2012MP}. A
straightforward extension of PL theory \cite{cuepre07} to the mixtures is the one-fluid approximation whereby one maps the mixture on to an effective
one-component HS system. A decoupling approximation is used in the PL approach in which the system is represented as the effective hard sphere of the same diameter
while any information about the geometry of the LC particles is included in the term of the factorized excluded volumes. In order to improve the PL treatment for
mixtures, a many-fluid (MF) approach has been proposed \cite{Malijevsky2008JCP} where each component in the mixtures are mapped on to the corresponding effective HS
system separately, thus LC mixtures are represented as mixtures of HS. Following the separate routes of Parsons and of Lee, two versions of many-fluid theories can be
developed: many-fluid Parsons (MFP) and many-fluid Lee (MFL) as alternatives for more accurate descriptions of LC mixtures. These many-fluid approaches have been
assessed for a mixture of hard Gaussian particles and it has been shown that MFP is superior to the PL and MFL methods at moderate and high densities
\cite{Malijevsky2008JCP}.

The focus of our current work is the isotropic-nematic phase behaviour of a HSC-HS mixture.
Previous reports of the ordering in the HSC-HS binary system have been presented including
direct simulation \cite{Koda1996JPSJ,Lago2004JMR,Peroukidis2010JMC}
and theoretical \cite{Roth2003EPL,Schmidt2003JCP,cuepre07} studies.
The work of Cuetos and co-workers is of particular relevance: the one-fluid PL
approach \cite{cuepre07} was used to study
the isotropic-nematic phase diagram of the HSC-HS system characterized by rods of various
lengths and diameters; comparisons where made with $NPT$ Monte Carlo simulations
\cite{Lago2004JMR} employed to
investigate the phase diagram and fluid structure of the mixtures. It is worth noting
that in the $NPT$ ensemble the system composition remains constant overall,
which will lead to an inadequate description of the phase boundary as one enters metastable states
which would otherwise phase separate into phases of distinct compositions.

The purpose of our current work is twofold.
First, the many-fluid Parsons theory is used to describe the HS-HSC system and comparisons
are made with the one-fluid PL approach.
It should be noted that in a one-component case both approaches reduce to the standard PL theory.
Second, we present new Monte Carlo simulation results for the mixture.
The local density (packing fraction), local composition,
and orientational distributions are determined during the simulations to estimate the
locations of the isotropic-nematic transitions of the mixture at various compositions
in order to make a proper test of the accuracy of the two theories.

\section{Theory of nematic phase in mixtures of hard particles}\label{chp5_Sec_Theory}

In this section, the main steps leading to a formulation with both one-fluid and
many-fluid theories are briefly recalled; further details can be found in Ref.
\citenum{Malijevsky2008JCP}.
Consider an $n$-component mixture system of $N$ unaxial (cylindrically
symmetrical)
hard anisotropic bodies in a volume $V$ at
a temperature $T$. The free energy functional of the system can be expressed as
a contribution from an ideal (entropy) term ($F^{\rm id}$)
and a residual (configurational) part ($F^{\rm res}$):

\begin{equation}
\label{chp5_Eq_MixBulk_FreeEn_tot}
\frac{\beta F}{V}=\frac{\beta F^{\rm id}}{V}+\frac{\beta F^{\rm res}}{V},
\end{equation}
where $\beta=1/(k_{\rm B}T)$ and $k_{\rm B}$ is the Boltzmann constant;
the temperature plays a trivial role in this case since only hard repulsive interactions between
particles are considered. The ideal free energy accounts for
the translational and orientational entropy and can be written as

\begin{equation}
\label{chp5_Eq_MixBulk_FreeEn_ideal}
\frac{\beta F^{\rm id}}{V}=\sum_{i=1}^{n} \rho_{i}\left\{\ln \left(\rho_{i} \mathcal{V}_{i}\right)-1+\sigma[f_{i}(\vec{\omega})]\right\},
\end{equation}

where $\rho_{i}=N_{i}/V$ ($N=\sum{}_{i=1}^{n}N_{i}$) is the number density of component $i$,
$\rho = N/V = \sum{}_{i=1}^{n} \rho_i$, and $\mathcal{V}_{i}$ is the de Broglie volume of each species,
incorporating the translational and rotational kinetic contributions of the ideal isotropic state.
With the introduction of single-particle orientational distribution function $f_{i}(\vec{\omega})$,
the orientational entropy term $\sigma[f_{i}]$ can be expressed as
an integration over all orientations $\vec{\omega}$ of a single particle:
\begin{equation}
\label{chp5_Eq_MixBulk_OrienEntropy}
\sigma[f_{i}(\vec{\omega})]=\int {\rm d}\vec{\omega} f_{i}(\vec{\omega}) \ln[4{\pi} f_{i}(\vec{\omega})] .
\end{equation}

For the residual part, Onsager's original expression \cite{Onsager1949} is equivalent to
truncating the virial expansion at second-virial level. At higher densities, however, the many-body correlations become progressively more and more important.
Following the Parsons approach \cite{Parsons1979}, we can include higher-body contributions in an approximate manner.
Assuming a pairwise additive hard interaction $u_{ij}(r_{kl},\vec{\omega}_{k},\vec{\omega}_{l})$
between particle $k$ of $i$-th component and particle $l$ of $j$-th component,
the pressure of a fluid mixture of $n$ components can be written in the virial form as \cite{GrayandGubbins1984}
\begin{equation}
\label{chp5_Eq_MixBulk_Pvirial}
P=\rho k_{\rm B}T-\frac{1}{6V}\Bigg\langle {\sum_{i=1}^{n}}{\sum_{j=1}^{n}}
{\sum_{k=1}^{N_{i}}}\sum_{l=1 \atop l\neq k}^{N_{j}} r_{kl}
\frac{\partial u_{ij}(\vec{r}_{kl},\vec{\omega_{k}},\vec{\omega_{l}})}
{\partial r_{kl}}\Bigg\rangle ,
\end{equation}
where $l\neq k$ is used to avoid self interactions and $\langle\cdots\rangle$
represents the ensemble average. In the canonical ensemble,
\begin{equation}
\label{chp5_Eq_MixBulk_Pvirial_CanonicalEnsemble}
P=\rho k_{\rm B} T-\frac{Z^{-1}}{6V} \prod_{i=1}^{n} \int {\rm d}{\vec{r}^{N_{i}}} \int {\rm d}{\vec{\omega}^{N_{i}}}
{\sum_{i=1}^{n}}{\sum_{j=1}^{n}}
{\sum_{k=1}^{N_{i}}}\sum_{l=1 \atop l\neq k}^{N_{j}} r_{kl}
\frac{\partial u_{ij}(\vec{r}_{kl},\vec{\omega_{k}},\vec{\omega_{l}})}
{\partial r_{kl}}\exp(-\beta U)
\end{equation}
where the configurational partition function $Z$ is defined as
\begin{equation}
\label{chp5_Eq_MixBulk_ConfigPartition}
Z=\prod_{i=1}^{n} \int {\rm d}{\vec{r}^{N_{i}}} \int {\rm d}{\vec{\omega}^{N_{i}}}
\exp(-\beta U(\vec{r}^{N_{i}},\vec{\omega}^{N_{i}})) ,
\end{equation}
and the total configurational energy is given by
\begin{equation}
\label{chp5_Eq_MixBulk_TotalConfigEn}
U(\vec{r}^{N_{i}},\vec{\omega}^{N_{i}})=\frac{1}{2}{\sum_{i=1}^{n}}{\sum_{j=1}^{n}}
{\sum_{k=1}^{N_{i}}}\sum_{l=1 \atop l\neq k}^{N_{j}} u_{ij}(\vec{r}_{kl},\vec{\omega_{k}},\vec{\omega_{l}}).
\end{equation}
The canonical pair distribution function is defined as
\begin{equation}
\label{chp5_Eq_MixBulk_PairDistFunc}
g_{ij}(\vec{r}_{12},\vec{\omega_{1}},\vec{\omega_{2}})=\frac{N_{i}(N_{j}-\delta_{ij})}
{\rho_{i}f_{i}(\vec{\omega}_{1})\rho_{j}f_{j}(\vec{\omega}_{2})}Z^{-1}
\int {\rm d}{\vec{r}^{N-2}} \int {\rm d}{\vec{\omega}^{N-2}} \exp(-\beta U(\vec{r}^{N_{i}},\vec{\omega}^{N_{i}})) ,
\end{equation}
where $\delta_{ij}$ is the Kronecker delta. On integrating Equation (\ref{chp5_Eq_MixBulk_Pvirial_CanonicalEnsemble}),
the expression for pressure can be written in a compact form:
\begin{eqnarray}
\label{chp5_Eq_MixBulk_Pressure}
P&=&\rho k_{\rm B} T-\frac{1}{6}\sum_{i=1}^{n}\sum_{j=1}^{n} \rho_{i}\rho_{j}
\int {\rm d}{\vec{r}_{12}} \int {\rm d}{\vec{\omega}_{1}} \int {\rm d}{\vec{\omega}_{2}}  \nonumber \\
&\times& r_{12} \frac{\partial u_{ij}(\vec{r}_{12},\vec{\omega_{1}},\vec{\omega_{2}})}
{\partial r_{12}} g_{ij}(\vec{r}_{12},\vec{\omega_{1}},\vec{\omega_{2}}) f_{i}(\vec{\omega}_{1})f_{j}(\vec{\omega}_{2}).
\end{eqnarray}

Following the Parsons approach, the interparticle separation $\vec{r}_{12}$ is
given in terms of the contact distance
$\sigma_{ij}(\vec{r}_{12},\vec{\omega_{1}},\vec{\omega_{2}})$ by defining a
scaled distance
$y_{ij}=r_{12}/\sigma_{ij}(\vec{r}_{12},\vec{\omega_{1}},\vec{\omega_{2}})$.
The scaled distance does not explicitly depend on the orientations of the two
particles and $y_{ij}=1$ corresponds to contact value.  Using the definition of
$y_{ij}$, the pair distribution function (cf. Equation
(\ref{chp5_Eq_MixBulk_PairDistFunc})) can be expressed as a function of scaled
distance $y_{ij}$, i.e., $g_{ij}=g_{ij}(y)$, which decouples the positional and orientational
dependencies. In this way, a complicated pair potential $u_{ij}$ is mapped onto
the spherically symmetrical hard-sphere potential:
\begin{eqnarray}
\label{chp5_Eq_MixBulk_PotentalScaled}
u_{ij}(\vec{r}_{12},\vec{\omega_{1}},\vec{\omega_{2}}) =
u_{ij}(y)=\begin{cases}
 \infty & y<1 \\
 0 & y\geq1,
\end{cases}
\end{eqnarray}
and the expression for the pressure becomes
\begin{eqnarray}
\label{chp5_Eq_MixBulk_Pressure2}
P&=&\rho k_{\rm B} T-\frac{1}{6}\sum_{i=1}^{n}\sum_{j=1}^{n} \rho_{i}\rho_{j}
\int {\rm d}y_{ij} y_{ij}^{3}\frac{{\rm d}u_{ij}}{{\rm d}y_{ij}} g_{ij}(y)  \nonumber \\
&{}&\times \int {\rm d}\hat{r}_{12} \int {\rm d}\vec{\omega}_{1} \int {\rm d}\vec{\omega}_{2} f_{i} (\vec{\omega}_{1})
f_{j} (\vec{\omega}_{2})\sigma_{ij}^{3} (\hat{r}_{12},\vec{\omega_{1}},\vec{\omega_{2}})  \nonumber \\
&=& \rho k_{\rm B} T-\frac{1}{2}\sum_{i=1}^{n}\sum_{j=1}^{n} \rho_{i}\rho_{j}
\int {\rm d}y_{ij} y_{ij}^{3}\frac{{\rm d}u_{ij}}{y_{ij}} g_{ij}(y)  \nonumber \\
&{}&\times \int {\rm d}\vec{\omega}_{1} \int {\rm d}\vec{\omega}_{2} f_{i} (\vec{\omega}_{1})
f_{j} (\vec{\omega}_{2}) V_{ij}^{\rm exc} (\vec{\omega_{1}},\vec{\omega_{2}})
\end{eqnarray}
where the excluded volume between a pair of particles is $V_{ij}^{\rm exc} (\vec{\omega_{1}},\vec{\omega_{2}})=
\frac{1}{3}\int {\rm d}\hat{r}_{12} \sigma_{ij}^{3} (\hat{r}_{12},\vec{\omega_{1}},\vec{\omega_{2}})$ and
$\hat{r}_{12}=\vec{r}_{12}/r_{12}$. The form of hard repulsive pair interaction is a step function
(cf. Equation (\ref{chp5_Eq_MixBulk_PotentalScaled})), thus $\beta{{\rm d}u_{ij}}/{{\rm d}y_{ij}}=-\exp(\beta u_{ij})\delta(y_{ij}-1)$ 
(for example, see Ref. \citenum{Brumby2011MP}).
Integrating over the scaled variable $y_{ij}$ and noting that $u_{1^{+}}(y)=0$ when $y=1$, we then obtain
\begin{equation}
\label{chp5_Eq_MixBulk_Pressure3}
P=\rho k_{\rm B} T+\frac{1}{2}\sum_{i=1}^{n}\sum_{j=1}^{n} \rho_{i}\rho_{j} g^{\rm HS}_{ij}(1^{+})
\int {\rm d}\vec{\omega}_{1} \int {\rm d}\vec{\omega}_{2} f_{i} (\vec{\omega}_{1}) f_{j} (\vec{\omega}_{2})
V_{ij}^{\rm exc} (\vec{\omega_{1}},\vec{\omega_{2}}),
\end{equation}
where $g_{ij}(1^{+}) \approx g^{\rm HS}_{ij}(1^{+})$ has been approximated as
the corresponding hard-sphere contact value of pair distribution function.

The residual free energy can then be obtained from the formal thermodynamic definition $(\partial F /\partial V)_{NT} = -P$
by integrating Equation (\ref{chp5_Eq_MixBulk_Pressure3}) over the volume:
\begin{equation}
\label{chp5_Eq_MixBulk_FreeEn_res}
\frac{\beta F^{\rm res}}{V}=\frac{1}{2}\sum_{i=1}^{n}\sum_{j=1}^{n} \rho_{i}\rho_{j} G_{ij}
\int {\rm d}\vec{\omega}_{1} \int {\rm d}\vec{\omega}_{2} f_{i} (\vec{\omega}_{1}) f_{j} (\vec{\omega}_{2})
V_{ij}^{\rm exc} (\vec{\omega_{1}},\vec{\omega_{2}}),
\end{equation}
where $G_{ij}=\rho^{-1} \int_{0}^{\rho}{\rm d}{\rho '} g^{\rm HS}_{ij}(1^{+})$.
Onsager's second-virial theory can be recovered with $G_{ij}=1$
(i.e., $g^{\rm HS}_{ij}(1^{+})=1$) corresponding to the low-density limit.

In this way, the theory of Parsons for a one-component fluid
can be reformulated to describe a $n$-component mixtures of anisotropic bodies.
As shown in Ref. \citenum{Malijevsky2008JCP}, the standard  ``one-fluid''
approach \cite{cuepre07} corresponds to $G_{ij}=G^{\rm PL}$, where
$G^{\rm PL}=\rho^{-1} \int_{0}^{\rho}{\rm d}{\rho '} g^{\rm HS}_{\rm CS}(1^{+})$,
given in terms of the Carnahan-Starling form of the radial distribution function at contact \cite{Carnahan1969,Hansen2006book},

\begin{equation}
\label{chp5_Eq_MixBulk_ContactVal_PL}
g^{\rm HS}_{\rm CS}(1^{+})=\frac{1-\eta/2}{(1-\eta)^3},  %Parsons-Lee approach
\end{equation}

\noindent with $\eta=\sum_{i=1}^{n}\rho_{i}V_{{\rm m},i}$, and $V_{{\rm m},i}$
is the volume of the $i$-th species. The PL residual free energy can then be expressed as

\begin{equation}
\label{chp5_Eq_MixBulk_FreeEn_PL}
\frac{\beta F^{\rm res,PL}}{V}=\frac{\rho^{2}}{8} \frac{4-3\eta}{(1-\eta)^3}\sum_{i=1}^{n}\sum_{j=1}^{n} x_{i}x_{j}
\int {\rm d}\vec{\omega}_{1} \int {\rm d}\vec{\omega}_{2} f_{i} (\vec{\omega}_{1}) f_{j} (\vec{\omega}_{2})
V_{ij}^{\rm exc} (\vec{\omega_{1}},\vec{\omega_{2}}) ,
\end{equation}

%Many-fluid approach
Alternatively, in developing the many-fluid theory proposed in Ref. \citenum{Malijevsky2008JCP}
one treats the size (volume) of each species individually, i.e.,
\begin{equation}
\label{chp5_Eq_MixBulk_MFP_Map}
V_{{\rm m},i}=V_{{\rm HS},i}=\frac{\pi}{6}{{\sigma}_{i}^3}, {\quad }i=1,2, \cdots,n.
\end{equation}
An expression for the contact value of the distribution function for hard-sphere mixture is given by Boublik \cite{Boublik1970JCP}:
\begin{equation}
\label{chp5_Eq_MixBulk_MFP_ContactVal_HSMix}
g_{ij,\rm B}^{\rm HS,Mix}(1^{+})=\frac{1}{1-\zeta_{3}}+\frac{3\zeta_{2}}{(1-\zeta_{3})^2}
\frac{\sigma_{ii}\sigma_{jj}}{\sigma_{ii}+\sigma_{jj}}+\frac{2\zeta_{2}^{2}}{(1-\zeta_{3})^3}
\frac{(\sigma_{ii}\sigma_{jj})^2}{(\sigma_{ii}+\sigma_{jj})^2}
\end{equation}
where the moments of the density are defined as $\zeta_{\alpha}=(\pi/6)\sum_{i=1}^{n}\rho_{i}\sigma_{ii}^{\alpha}, \quad \alpha=0,1,2,3$.
Combining Equations (\ref{chp5_Eq_MixBulk_FreeEn_res}) and (\ref{chp5_Eq_MixBulk_MFP_ContactVal_HSMix})
and noting the separate definition of $G_{ij}$ for each $i-j$ pair,
one obtains the many-fluid Parsons (MFP) form of the residual free energy $F^{\rm res,MFP}$.

In the one-component limit the contact value of the radial distribution function of
the HS mixture (Equation (\ref{chp5_Eq_MixBulk_MFP_ContactVal_HSMix}))
reduces to the Carnahan-Starling expression (cf. Equation(\ref{chp5_Eq_MixBulk_ContactVal_PL})).
Thereby, the MFP approach yields same descriptions as
PL treatment for the pure-component systems. In the standard extension of the PL theory to mixtures
one therefore adopts a van der Waals one-fluid (VDW1) approximation using an equivalent hard-sphere system
with the effective diameter given by the VDW1 mixing rule to represent the anisotropic mixtures.
In contrast to the PL approach, each component is represented as a separate effective hard-sphere component,
so that the excluded volume between a pair of $i$-th component and $j$-th component
is weighted by the corresponding contact value of the HS mixture, $g_{ij}^{\rm HS}$ .
The equilibrium free energy of the system is determined from a functional variation with
respect to the orientational distribution function $f_{i}(\vec{\omega})$ of each component
which leads to an integral equation for $f_{i}(\vec{\omega})$. The set of integral equations
are solved numerically using an iterative procedure, 
details of which can be found in Ref. \citenum{Malijevsky2008JCP}.

\begin{figure}[ht]
\centering
 \includegraphics[angle=0,scale=0.50]{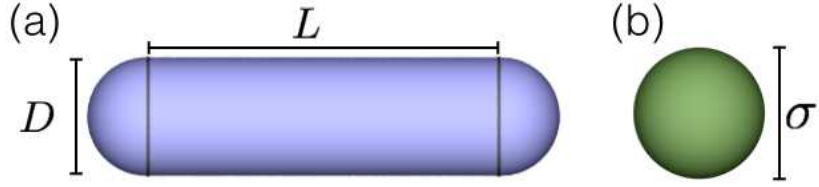}
  \caption{\label{fig1}The hard-core models: (a) hard spherocylinder (HSC) of length $L$ and diameter $D$;
  and (b) hard sphere (HS) of diameter $\sigma$. In the current study, the length of the HSC is fixed to $L=5D$ and
  the diameter of the HS is the same value as that of the HSC, i.e., $\sigma=D$.}\label{chp5_fig_MixBulk_model}
\end{figure}

In this work, we assess the adequacy of many-body theories such as the MFP for
a binary mixture of hard spheres and hard spherocylinders.
The models are depicted in Figure \ref{chp5_fig_MixBulk_model}:
the aspect ratio of the HSC is $L/D=5$ and
the diameter of the HS is taken to be the same as the diameter of the HSC, i.e., $\sigma=D$.

The excluded volumes corresponding to the HSC-HSC, HSC-HS and HS-HS interactions are given as
\begin{eqnarray}
\label{chp5_Eq_MixBulk_Vexc}
V_{\rm HSC-HSC}^{\rm exc}&=&\frac{4}{3}\pi {D^3}+2\pi L{D^2}+2L^{2}D|\sin \gamma| \nonumber \\
V_{\rm HSC-HS}^{\rm exc}&=&\frac{\pi}{6}{(D+\sigma)^3}+\frac{\pi}{4}L{(D+\sigma)^2}  \nonumber \\
V_{\rm HS-HS}^{\rm exc}&=&\frac{4}{3}\pi \sigma^3.
\end{eqnarray}
where $\gamma=\arccos(\vec{\omega}_{1}\cdot\vec{\omega}_{2})$ is the
relative orientation of the two HSC particles.
The total excluded volume of the mixture is $V^{\rm exc}=x_{\rm HSC}^{2}V_{\rm HSC-HSC}^{\rm exc}+
2x_{\rm HSC}x_{\rm HS}V_{\rm HSC-HS}^{\rm exc}+x_{\rm HS}^{2}V_{\rm HS-HS}^{\rm exc}$ where $x_{\rm HS}$ and $x_{\rm HSC}$
are mole fractions of HS and HSC species, respectively. Since the HSC particles are the anisotropic component in the system,
$f(\vec{\omega})$ is used to describe the orientation distribution of the HSC rods
which is related to the nematic order parameter $S_{2}$ of the system through
\begin{equation}
\label{chp5_Eq_MixBulk_S2}
S_{2}=\int {\rm d} \vec{\omega} f(\vec{\omega}) \left(1-\frac{3}{2}\sin^{2}\gamma\right).
\end{equation}
In particular, $S_{2}=0$ corresponds to the isotropic state and $S_{2}=1$ for a perfectly-aligned nematic phase.

\section{Monte Carlo simulation of phase coexistence in mixtures of hard spheres and hard spherocylinders} \label{chp5_Sec_Simu}

There are two common approaches to studying fluid-phase separation by molecular simulation. Within the direct procedure the two coexisting phases are treated
simultaneously in the presence of an interface with the usually periodic boundary conditions \cite{Allen1989simuBook,Frenkel2001simuBook}. The stabilization of a
fluid interface corresponding to a system with a nonuniform density within a single
simulation box is straightforward to implement with either molecular dynamics (MD)
or Monte Carlo (MC) techniques. This was first demonstrated by Croxton and Ferrier \cite{Croxton1971JP} who performed MD simulations of the vapor-liquid interface of
a Lennard-Jones system in two dimensions, and shortly afterwards by Leamy et al. \cite{Leamy1973PRL} who stabilized the interface of a three-dimensional lattice gas
(Ising model) by MC simulations. For a system which is sufficiently large (in the direction normal to the interface) one can simultaneously examine the bulk
properties, in the central region of the coexisting phases as well as the interfacial properties.

The direct molecular simulation of the isotropic-nematic (I-N) phase transition in mixtures
of hard spheres and hard spherocylinders is particularly challenging
because of the very low interfacial tension between the two phases;
for example, the I-N interfacial tension of a hard-core system of for thin disc-like particles
has be estimated to be a few tenth of $k_{\rm B} T$ in units of the particle's area \cite{vdBeek2006PRL}.
As a consequence there is a very low energetic penalty associated with the deformation of the interface in such systems
leading to large interfacial fluctuations; moreover, in the absence of an external field there is no resistance to the translation of a planar interface.
The location of the bulk coexistence regions and the determination of the density and compositional profiles
becomes a difficult task as a result. In order to break the symmetry of the system and reduce
the effect of the interfacial fluctuations one can introduce an external field by
placing the system within structureless hard walls; this corresponds to removing the periodicity
in one dimension (say the $z$ direction). An issue with this type of approach is that large systems
have to be considered in order to study the true bulk phase behaviour and avoid capillary effects.
By keeping the separation between the walls large compared to the dimensions of the particles,
one can simulate the phase coexistence in the hard-core HS-HSC mixtures with minimal effect from the hard surfaces.

Alternatively, the phase behaviour can be simulated using a popular Gibbs ensemble  \cite{Allen1989simuBook,Frenkel2001simuBook} in which the coexisting phases are
retained in separate boxes and coupled volume changes and particles exchanges between the boxes are undertaken to meet the requirements of mechanical and chemical
equilibria. However, in the case of hard anisometric particles, the acceptance ratio for the insertion of anisotropic particles will be very low, particularly at the
high densities of the dense anisotropic phases of interest, requiring an impracticably large number of trial insertions for a proper equilibration of the system
\cite{Dijkstra1997PRE}. A conventional simulation of the system within a single box will partially solve the problem since trial insertions of the particles are no
longer required. There is however a complication with the simulation of bulk phase equilibria of mixtures with a single simulation cell: though the phase transition
between the various states can be traced as for a pure component system, the overall composition remains fixed preventing the fractionation of the different species
in the various phases.

In view of the aforementioned issues, we employ a less conventional $NP_nAT$
ensemble within a single cell where the component $P_n$ of the pressure tensor
normal to the interface is kept constant, so that the condition of mechanical
equilibria is satisfied within the entire simulation cell
\cite{deMiguel2006MP,Brumby2011MP,Brumby2010PHD}. The advantage of simulating
the phase separation of mixtures by simultaneously considering the coexisting
phases and the interface in a single cell is that this will allow for
inhomogeneities in both the density and the composition of the system. By
introducing an external field such as a hard surface one is able to examine both
the bulk and interfacial regions of mixtures of hard core particles without
constraining the density or composition of the individual bulk phases.

\begin{figure}[ht]
\centering
\includegraphics[angle=0,scale=0.45]{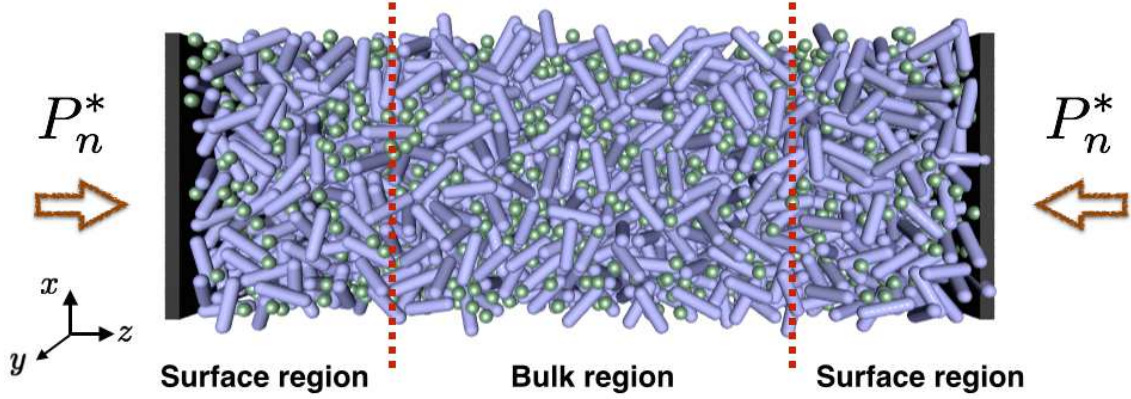}\\
\caption{The $NP_nAT$-MC simulation cell: hard walls are placed along the z axis and
the cell is divided into three large bins: two surface regions close to the hard walls and
a bulk region in the central part of the cell. A fixed normal pressure is imposed and
the dimension of the system is allowed to fluctuate in the z direction.
In this example a mixture system of hard spherocylinders (purple rods)
and hard spheres (green spheres) is depicted.}\label{chp5_fig_MixBulk_SimulationBox}
\end{figure}

We perform constant normal-pressure Monte Carlo simulation ($NP_nAT$-MC) for a
system of $N_{\rm HSC}=1482$ HSC particles of the aspect ratio $L/D=5$ where the
number of hard spheres is varied depending on mole fraction of the binary
mixture $x_{\rm HS,tot}=N_{\rm HS}/(N_{\rm HS}+N_{\rm HSC})$. In this system,
the intermolecular potential between any two particles is restricted to a pure
repulsion. As shown in Figure \ref{chp5_fig_MixBulk_SimulationBox}, the
simulation cell is a rectangular box of dimension $L_{x}=L_{y}=25D$
(corresponding to a fixed surface area $A$ in the $x$-$y$ plane of $A=625D^2$) and
$L_{z}$ varies according to the value set for $P_n$.  The parallel hard walls
are positioned at $z=0$ and $z=L_{z}$ and standard periodic boundary conditions
are applied in $x$ and $y$ directions. Since a fixed normal pressure is imposed
along the $z$ axis, the system volume in our $NP_nAT$-MC simulation is allowed
to fluctuate by scaling the length of the $z$ axis which moves the walls closer
together or farther apart, while the system dimensions of the $x$ and $y$ axes
and the $x$-$y$ surface area are kept fixed.

The $NP_nAT$-MC simulation of the HSC-HS mixture is performed for
$5\times10^{6}$ cycles to equilibrate the system and $5$ to $8\times10^{6}$
cycles to obtain the average properties. Each MC cycles consist of $N=N_{\rm
HS}+N_{\rm HSC}$ attempts to displace and rotate (in the case of a HSC particle)
randomly chosen particles and one trial volume change corresponding to a
contraction or extension in the $z$ direction. The breaking of symmetry caused
by the hard walls leads to inhomogeneous positional, orientational, and
compositional distributions of the system along the $z$-axis, so that the
thermodynamic and structural properties have to be determined locally. Smooth
density and composition profiles are required to identify the uniform region in
the centre of the box which correspond to the bulk phase. In order to evaluate
the packing fraction $\eta_{i}(z)$, composition $x_{\rm HS}(z)$, and order
parameter $S_{2}(z)$ profiles, the simulation box is divided into several bins
of equal width $\delta z$ in the $z$ direction; $n_{\rm bin}=200$ bins are used to
calculate the packing fraction profile $\eta_{i}(z)=\rho_{i}(z)V_{{\rm m},i},
i={\rm HS, HSC}$, where the number density profile of the component $i$ is
obtained from
\begin{equation}
\rho_{i}(z)=\frac{\langle N_{i}(z) \rangle}{L_{x}L_{y}\delta z} \qquad i={\rm HS, HSC}
\end{equation}
and the local composition is then obtained as $x_{\rm HS}(z)=\rho_{\rm HS}(z)/(\rho_{\rm HS}(z)+\rho_{\rm HSC}(z))$.
The packing fraction $\eta_{\rm b}$ and composition $x_{\rm HS,b}$ of the bulk phase is then determined to be the values
corresponding to the uniform regions of the packing fraction and composition profiles in the centre of the simulation cell.

The nematic order parameter profile $S_{2}$ is obtained by determining
the local nematic order parameter tensor $\mathbf{Q}(j)$ in each bin $j$:
\begin{equation}
{
\mathbf{Q}(j)=\left<\frac{1}{2N_{{\rm HSC},j}} \sum_{i=1}^{N_{{\rm HSC},j}}
\left(3\vec{\omega}_{i} \otimes  \vec{\omega}_{i}-\mathbf{I}\right) \right>
}
\label{chp5_Eq_MixBulk_OrderPara_Tensor}
\end{equation}
where $N_{{\rm HSC},j}$ is the number of HSC particles in the $j$-th bin and $\mathbf{I}$
is the unit tensor.
On diagonalising the tensor $\mathbf{Q}(j)$, three eigenvalues can be obtained and
the largest eigenvalue defines the local nematic order parameter $S_{2}(z)$ of the $j$-th bin.
Special care is required in calculating the order parameter profile because of finite-size effects.
It has been shown by Eppenga and Frenkel \cite{Eppenga1984} that the value of
the nematic order parameter depends on the number of particles considered and
the error in local order parameter is $\sim 1/(\sqrt{N_{\rm HSC}/n_{\rm bin}})$.
If we use $n_{\rm bin}=200$ which is the same number of bins used for density profile,
there are on average only 7$(\approx1482/200)$ rods in each bin and the expected error
in $S_{2}(z)$ of $\sim 0.367$ is large. Richter and Gruhn \cite{Richter2006JCP} have employed
a methodology to correct for finite-size effects by introducing a function which
bridges $S_{2,N_{\rm HSC} \rightarrow \infty}(z)$ and $S_{2,N_{\rm HSC}}(z)$ by correlating the simulation data.
A more direct way \cite{Brumby2010PHD} to reduce the error in the local nematic order parameter is
to examine a larger system; for example, in the case of a system of 14820 rods (an order of magnitude larger than the system studied here)
and 200 bins reduces the error in $S_{2}(z)$ to $\sim 0.11$.
However, the simulation of a system of this size is very computationally expensive.
As in our current work the focus is the bulk phase behaviour not the interfacial region, we divide the system into
3 large bins: 2 surface regions adjacent to the hard walls and the bulk region (cf. Figure \ref{chp5_fig_MixBulk_SimulationBox}).
With $n_{\rm bin}=3$ system-size error in $S_{2}(z)$ decreases to $\sim 0.04$ where there
are now an average of $\sim 500$ rods in each bin.
The value of $S_{2}(z)$ corresponding to the central region is
then taken to represent the nematic order parameter of the bulk phase $S_{2,\rm b}$.
The reduced normal pressure is defined as $P^{*}_{n}=P_nD/k_{\rm B}T$.

\section{Results and discussion} \label{chp5_Sec_result}

\subsection{Pure hard spherocylinders}

\begin{figure}[ht]
\centering
\includegraphics[angle=0,scale=0.45]{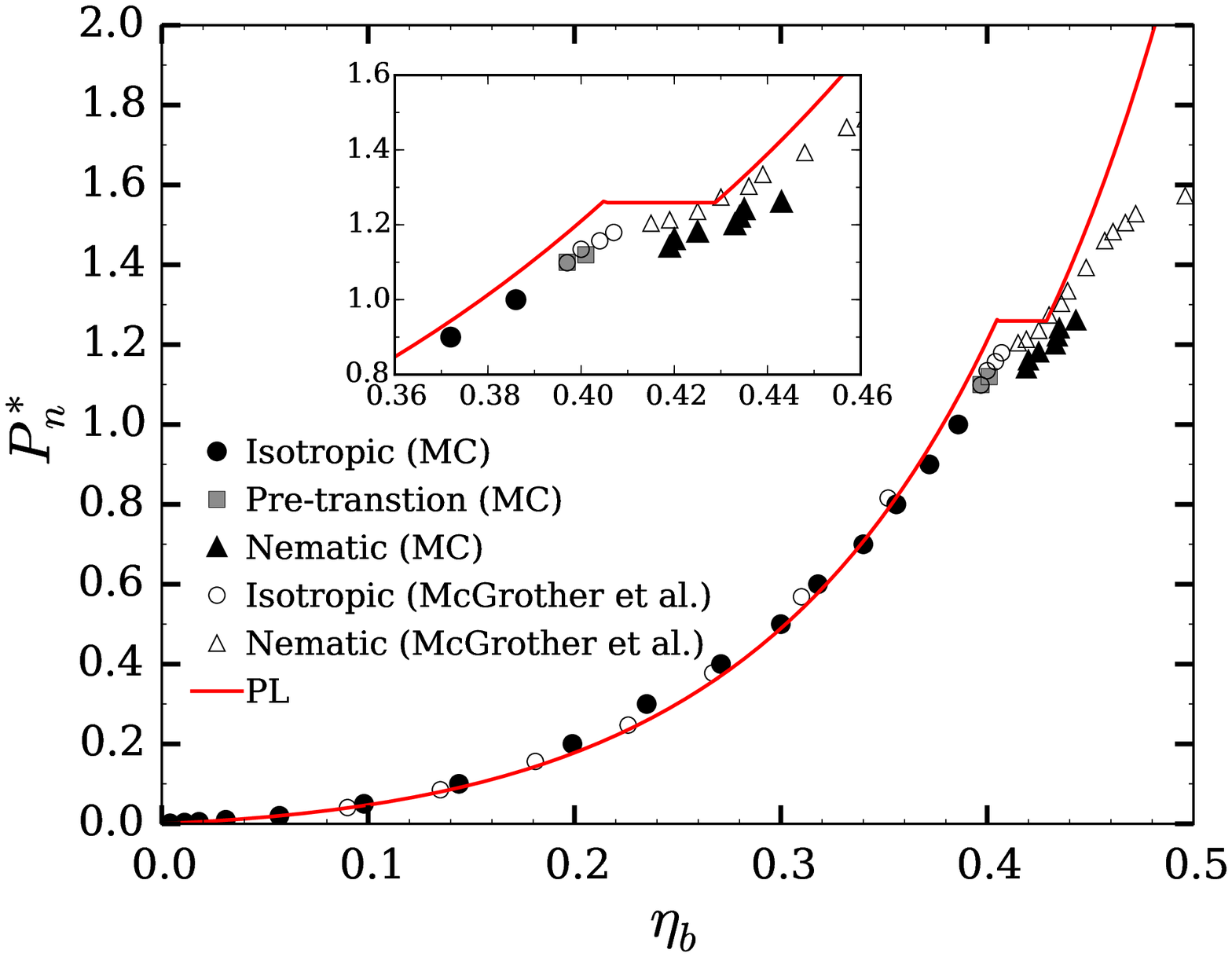}\\
 \caption
 {The isotropic-nematic phase behaviour of hard-spherocylinder (HSC) rods with
 an aspect ratio of $L/D=5$.  The simulation data obtained for the system of
 $N_{HSC}=1482$ particles contained between parallel hard surfaces using our
 $NP_nAT$-MC approach (filled symbols), where $P^{*}_n=P_nD^{3}/(k_{\rm B}T)$ is
 the dimensionless normal pressure and $\eta_{\rm b}=\rho_{\rm b} V_{\rm HSC}$
 is the bulk value of the packing fraction in the central region of the cell,
 are compared with the corresponding simulation data obtained  by McGrother {\it
 et al.} \cite{McGrother1996} for the fully periodic system using the
 conventional $NPT$-MC approach (open symbols), where now $P^*=P/(k_{\rm B} T)$
 and $\eta=\rho V_{HSC}$ are the values for the homogeneous system.  The circles
 corresponds to the isotropic states, the triangles to the nematic state, and
 the squares to the pre-transitions states.   The curve is
 the theoretical predictions obtained with the PL theory.  An enlargement of the
 isotropic-nematic transition region is shown in the
 inset.}\label{chp5_fig_MixBulk_PureRods_PrsEta}
\end{figure}

Prior to demonstrating our results for mixtures of HS and HSC particles, it is instructive to begin
by studying a system of pure HSC particles with aspect ratio of $L/D=5$.
As is well known, the simple HSC model of a mesogen exhibits isotropic, nematic,
smectic-A, and solid phases as the density of the system is increased
\cite{Stroobants1986PRL,Stroobants1987PRA,Frenkel1987JPC,Frenkel1988Nature,
Frenkel1988JPC,Veerman1991PRA,McGrother1996,Bolhuis1997JCP}.
As an preliminary assessment we demonstrate that the bulk isotropic-nematic transition
for the $L/D=5$ HSC system contained between well separated parallel hard surfaces
simulated using our $NP_nAT$-MC methodology is essentially unaffected by the external field.
The bulk phase behaviour for the homogeneous system obtained using conventional constant pressure
$NPT$-MC with full three dimensional periodic boundary conditions \cite{McGrother1996}
is compared with the corresponding data obtained using the constant normal-pressure
$NP_nAT$-MC methodology for the system between parallel hard surfaces in Figure \ref{chp5_fig_MixBulk_PureRods_PrsEta}
and corresponding simulation data is reported in Table \ref{chp5_table_SimuData_PureHSC}.

\begin{table}[htbp]
\caption{Constant normal-pressure MC ($NP_nAT$-MC) simulation results for bulk
isotropic-nematic phase behaviour of $N_{HSC}=1482$ hard spherocylinders with an aspect ratio $L/D=5$.
The reduced normal pressure $ P_n^{*}$ is set in the simulation and corresponding
bulk values of packing fraction $\eta_{\rm b}$, nematic order parameter $S_{2,\rm
b}$, and box length $L_z$
are obtained as configurational averages. The isotropic phase is denoted by Iso,
the nematic by Nem, and the pre-transitional states by Pre.}
\centering
\begin{tabular*}{1.00\textwidth}{@{\extracolsep{\fill}} ccccc}
\toprule
$ P_n^{*}$  & $\eta_{\rm b}$  & $S_{2,\rm b}$ & $ L_{z}/D$  & Phase \\
\colrule
0.001   &   0.004  &  0.042 & 899.85 & Iso  \\
0.003   &   0.011  &  0.042 & 881.84 & Iso  \\
0.005   &   0.018  &  0.042 & 562.67 & Iso  \\
0.01    &   0.031  &  0.042 & 327.44 & Iso  \\
0.02    &   0.057  &  0.043 & 183.67 & Iso  \\
0.05    &   0.098  &  0.045 & 105.10 & Iso  \\
0.10    &   0.144  &  0.046 & 71.12  & Iso  \\
0.20    &   0.199  &  0.049 & 51.17  & Iso  \\
0.30    &   0.235  &  0.053 & 43.39  & Iso  \\
0.40    &   0.271  &  0.056 & 38.84  & Iso  \\
0.50    &   0.300  &  0.067 & 35.12  & Iso  \\
0.60    &   0.318  &  0.072 & 32.81  & Iso  \\
0.70    &   0.340  &  0.074 & 30.85  & Iso  \\
0.80    &   0.356  &  0.085 & 29.23  & Iso  \\
0.90    &   0.372  &  0.099 & 27.68  & Iso  \\
1.00    &   0.386  &  0.116 & 26.75  & Iso  \\
1.10    &   0.397  &  0.371 & 25.40  & Pre  \\
1.12    &   0.401  &  0.456 & 25.18  & Pre  \\
1.14    &   0.419  &  0.553 & 24.88  & Nem  \\
1.16    &   0.420  &  0.555 & 24.82  & Nem  \\
1.18    &   0.425  &  0.570 & 24.61  & Nem  \\
1.20    &   0.433  &  0.619 & 24.22  & Nem  \\
1.22    &   0.434  &  0.692 & 23.95  & Nem  \\
1.24    &   0.435  &  0.727 & 23.73  & Nem  \\
1.26    &   0.443  &  0.739 & 23.57  & Nem  \\
\botrule
\label{chp5_table_SimuData_PureHSC}
\end{tabular*}
\end{table}

\begin{figure}[ht]
\centering
\includegraphics[angle=0,scale=0.45]{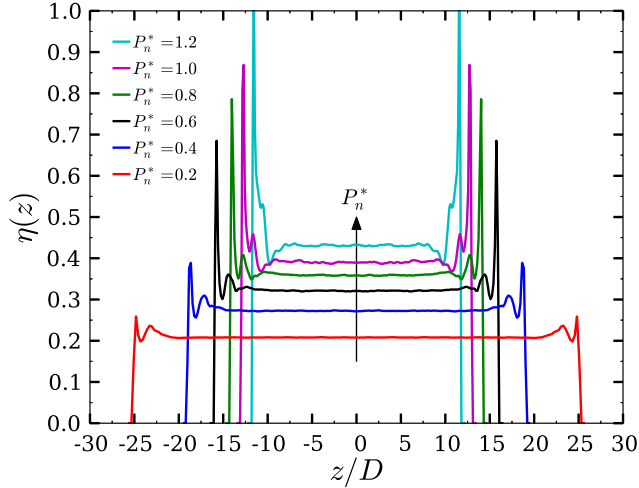}\\
\caption{Packing fraction profile $\eta(z)$ for $N_{HSC}=1482$ pure hard-spherocylinder
(HSC) rods with normal
pressure $P^{*}_{n}=0.2$ to $1.2$
obtained using $NP_nAT$-MC. Two hard walls are positioned at $z=0$ and $z=L_{z}$,
where $L_{z}$ is the length of the $z$ axis.}
\label{chp5_fig_Bulk_PureRod_Eta_Profile}
\end{figure}

The predictions of the isotropic and nematic branches with the MFP theory \cite{Malijevsky2008JCP}
is also shown in Figure \ref{chp5_fig_MixBulk_PureRods_PrsEta} which reduces to the commonly employed
PL theory \cite{Parsons1979,Lee1987,Lee1988} for one-component system: the theory is seen to provide a reasonably quantitative description
of the isotropic and nematic branches of the equation of state and the position of the isotropic-nematic transition.
We can infer that the results obtained for the HSC particles contained
between the parallel hard surfaces are fully consistent with
those of the fully periodic homogeneous system, confirming that at least for this system
size and geometry of the simulation cell the presence of hard surfaces
only has a small (stabilizing) effect on the isotropic-nematic transition;
in this case the separation between the surfaces, which defines the dimension the box in the $z$ direction,
ranges from $L_{z}\sim 899.85D$ at the lowest density ($P_n^{*}=0.001$)
studied to $L_{z}\sim 23.57D$ for the highest density ($P_n^{*}=1.26$) nematic state.
Example of the packing fraction (density) profiles $\eta(z)$ obtained for a low-density isotropic state,
an intermediate-density nematic state, and a moderately high-density nematic state are displayed in Figure \ref{chp5_fig_Bulk_PureRod_Eta_Profile};
the flat part of the profiles correspond to the homogenous bulk phases in the central region of
the simulation cell which allows one to determine the equilibrium bulk density $\eta_b$ with confidence.

As in other studies of confined liquid-crystalline systems
\cite{Dijkstra2001PRE,mulms03,Webster2003PRE,dijprl04,rulljccp07}, 
significant structure is also apparent close to the surfaces; a full analysis of the surface effects
such as wetting, de-wetting, surface nematization, and adsorption will be left for future
work.

\begin{figure}[ht]
\centering
\includegraphics[angle=0,scale=0.45]{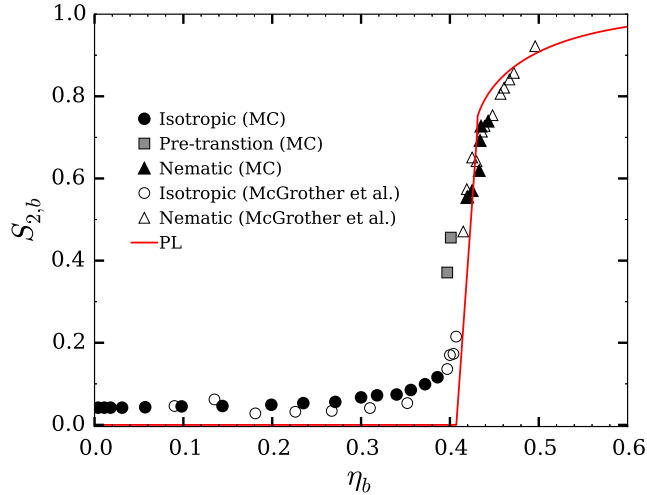}\\
 \caption
 {The density dependence of the bulk nematic order parameter $S_{2,\rm b}$
 for hard-spherocylinder (HSC) rods with aspect ratio $L/D=5$.
 The simulation data obtained for the system of $N=1482$ particles contained
 between parallel hard surfaces using our $NP_nAT$-MC approach (filled symbols),
 where $S_{2,\rm b}$ is the bulk value of the nematic order parameter
 and $\eta_{\rm b}=\rho_{\rm b} V_{\rm HSC}$ is the bulk value of the packing fraction
 in the central region of the cell, are compared with the corresponding simulation data
 obtained by McGrother {\it et al.} \cite{McGrother1996} for the fully periodic system using the conventional $NPT$-MC approach
 (open symbols),
 where now $S_{2}$ and $\eta=\rho V_{HSC}$ are the values for the homogeneous system.
 The circles correspond to the isotropic states, the triangles to the nematic states,
 and the squares the pre-transitional 
 states in the vicinity of isotropic-nematic transition.
 The curve is the theoretical prediction obtained with the PL theory.}\label{chp5_fig_MixBulk_PureRod_S2Eta}
\end{figure}

\begin{figure}[htbp]
\centering
\includegraphics[angle=0,scale=0.45]{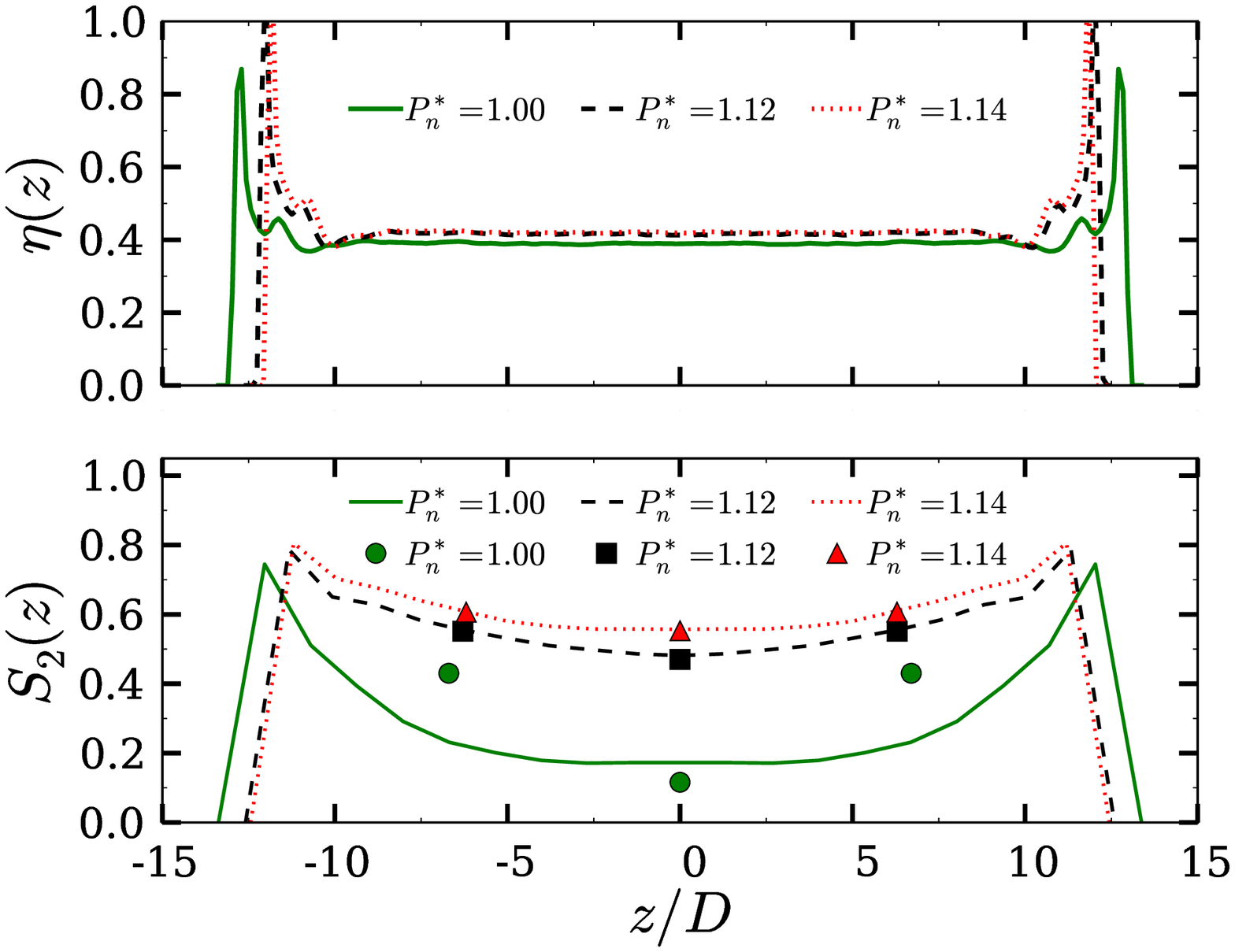}\\
\caption{\label{chp5_fig_MixBulk_PureRod_Profile} The nematic order parameter $S_2(z)$ (bottom panel) and packing fraction $\eta(z)$
(top panel) profiles
for hard-spherocylinder (HSC) rods with an aspect ratio of $L/D=5$ obtained by simulating
the system of $N=1482$ particles between parallel hard surfaces placed along the $z$ direction.
Isotropic ($P_n^{*}=1.00$), pre-transitional ($P_n^{*}=1.12$), and nematic ($P_n^{*}=1.14$) states
in the vicinity of the isotropic-nematic transition are examined;
the thermodynamic state is characterized by the value of the normal pressure tensor,
$P_n^{*}=P_n D^3/(k_{\rm B}T)$. In the case of $S_{2}(z)$ the profiles
are constructed from histograms with $n_{\rm bin}=3$ (symbols) and $n_{\rm bin}=20$
(curves) to assess possible system size effects.}
\end{figure}

\begin{figure}[htbp]
\centering
\includegraphics[scale=0.4]{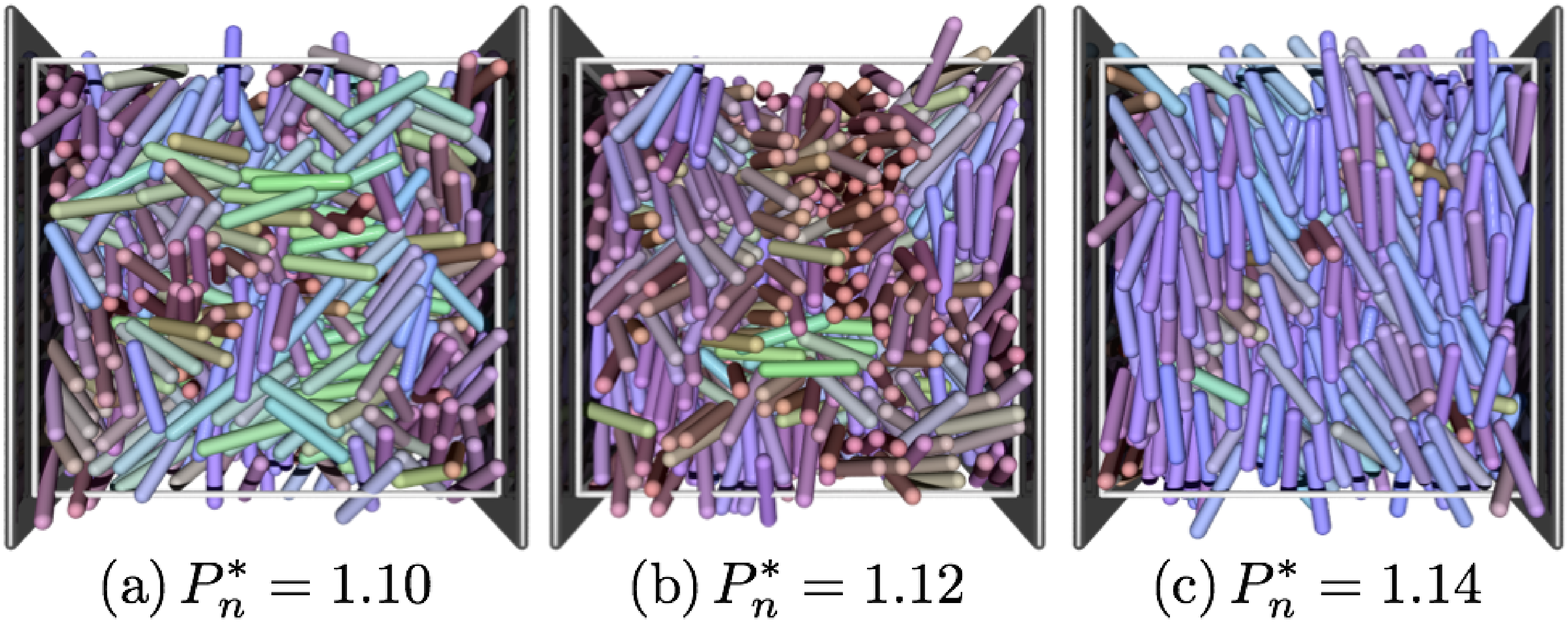}\\
\caption{Snapshots of typical configurations of $N_{HSC}=1482$ hard-spherocylinder (HSC) rods
with an aspect ratio of $L/D=5$ obtained by simulating the system between
parallel hard surfaces places along the $z$ axis.
(a) Isotropic ($P_n^{*}=1.00$), (b) pre-transitional ($P_n^{*}=1.12$),
and (c) nematic ($P_n^{*}=1.14$) states in the vicinity of the isotropic-nematic transition are examined
(see the Caption of Figure \ref{chp5_fig_MixBulk_PureRod_Profile} for further details).
The colours quantify the orientation of the rod-like particles relative to the frame of the simulation box.}\label{chp5_fig_MixBulk_PureRod_Snapshot}
\end{figure}

In order to estimate the location of the bulk isotropic-nematic bulk phase transition, we examine the
density dependence of the nematic order parameter in Figure \ref{chp5_fig_MixBulk_PureRod_S2Eta}.
The order parameter of a finite system $S_{2,N_{\rm HSC}}(z)$ converges quickly to
the limiting bulk thermodynamic value $S_{2,N_{\rm HSC} \rightarrow \infty}(z)$ for states with intermediate
to high orientational order ($S_{2,N_{\rm HSC}} \gtrsim 0.5$).
In Figure \ref{chp5_fig_MixBulk_PureRod_Profile} we display the order parameter profiles
for states of low to moderate orientational order
($0.1 \lesssim S_{2,N_{\rm HSC}}\lesssim 0.5$) in the close vicinity of the isotopic-nematic transition.
Snapshots of typical configurations of these two states are shown in Figure \ref{chp5_fig_MixBulk_PureRod_Snapshot}:
the low-density configuration corresponds to an bulk isotropic state,
the intermediate-density configuration to a pre-transitional
 state,
while the high-density configuration has clearly undergone a transition to a bulk nematic phase.
The pre-transition states, assumed here to correspond to nematic order parameters in the
range $0.3 \lesssim S_{2,N_{\rm HSC}}\lesssim 0.5$, are due to system size effects and can
also be
exacerbated by the confinement and potentially very slow relaxation order processes
in the form of slow nucleation kinetics.

The discrepancy between the values of the nematic order parameter obtained for profiles with $n_{\rm bin}=3$
and $n_{\rm bin}=20$ histogram bins in the z direction is very small so that the size effects are expected to be small in this case.
An isotropic bulk phase region is clearly found in the case of the state with a pressure of $P_n^{*}=1.00$;
in this case the orientational order in the bulk region is found to be low corresponding to an
nematic order parameter of $S_{2,b}=0.116$.
The order parameter profile for the pre-transitional state with a normal pressure of $P_n^{*}=1.12$
exhibits a  curve with no uniform bulk region;
the average of the nematic order parameter of $S_{2,\rm b}=0.456$
obtained in the central part of the cell does not therefore represent that of a true bulk phase.
In the case of the denser state corresponding to $P_n^{*}=1.14$,
the value of the bulk nematic order parameter $S_{2,\rm b}=0.553$ obtained as an average over $n_{\rm bin}=3$ bins
is essentially equivalent to the value of $S_{2,\rm b}=0.556$
obtained with $n_{\rm bin}=20$ bins in the homogeneous central region of the simulation cell.
The difference in the orientational ordering for the states corresponding to normal pressures
of $P_n^{*}=1.12$ and $P_n^{*}=1.14$ is also apparent from Figure \ref{chp5_fig_MixBulk_PureRod_Snapshot}
where the orientations of HSC rods have been colour coded to aid the visualization.
Small clusters of nematic domains are seen for the pre-transitional states
(Fig \ref{chp5_fig_MixBulk_PureRod_Profile}),
which lead to slightly larger values of the nematic order parameter $S_{2,b}$

Clearly, the equilibrium state at the pressure of $P_n^{*}=1.14$ corresponding to
a bulk density of $\eta_b=0.419$ and nematic order parameter of $S_{2,\rm b}=0.553$
can be taken to correspond to the lowest-density nematic state,
while the state at the slightly lower pressure of $P_n^{*}=1.12$ is seen to
exhibit some small nematic clusters which would correspond to a pre-transitional
 state;
large region with random orientations corresponding to
a bulk isotropic liquid can be seen in the case of the system with $P_n^{*}=1.00$.
Our estimate of the isotropic-nematic transition for the $L/D=5$ HSC system from
an analysis of this data is $\eta_{\rm b,iso}=0.386$, $\eta_{\rm b,nem}=0.419$ for the
bulk coexisting phases
which are in good agreement with the corresponding results estimated from
conventional $NPT$-MC simulation with full three-dimensional periodic boundary conditions
($\eta_{\rm iso}=0.407,\eta_{\rm nem}=0.415$)\cite{McGrother1996};
the coexistence pressure is arithmetic average between the normal pressures corresponding to
the highest-density isotropic state and the lowest-density nematic state, $P_n^{*}=1.07$,
which is slightly lower than that estimated for the system with full periodicity ($P_n^{*}=1.19$)\cite{McGrother1996}.
A determination of the free energy and chemical potential of the system would allow one
to get a more precise estimate of the position of the phase transition, but this is beyond the scope of the current study.
A slight stabilization of the isotropic-nematic transition (corresponding to a lowering the transition packing fraction by 1 to
$2\%$) is therefore found in our systems of HSC rods placed between two parallel hard walls.
One would certainly expect surface induced nematization to stabilize the transition to a bulk nematic phase.
These surface effects are however not the focus of the current study and will be discussed in detail in subsequent studies.
When the two confining walls are well separated the effect on the bulk isotropic-nematic transition is inappreciable.
However, the effects of confinement will become increasingly more important as the surfaces are brought closer together, i.e., at higher pressures and/or for small system sizes.
For the systems studied in our current work the two hard walls in our simulation cells are far enough apart
(even for the highest density states) so that the effect of the surfaces on the bulk phase transition is small.

\subsection{HSC-HS mixture }

We now turn our attention to a binary mixture of HS and HSC particles
characterized by overall HS composition $x_{\rm HS}$ ranging from 0.05 to 0.20.
As with the pure component system the mixture is contained between parallel hard
structureless surfaces and the isotropic-nematic phase transition is simulated
using the $NP_nAT$-MC method ({\it cf.} Section \ref{chp5_Sec_Simu}). As well as
being of different density the coexisting isotropic and nematic phases will
exhibit a fractionation of the two components, with an accumulation of the the
rod-like particles in the orientationally ordered nematic phase. The fluid phase
separation in hard-core system of this type is an entropy driven process. This
is not to be confused with the depletion driven phase behaviour exhibited by
anisometric colloidal particles on addition of polymer where the polymer induces
an effective attractive interaction (depletion force) between the colloids, that
would otherwise only interact in a purely repulsive fashion, giving rise to a
van der Waals like ``vapour-liquid'' transition (see for example references
[\citenum{Boluis1994JCP,Warren1994JPFr,Sear1995JCP,Vilegenthart1999JCP,Chen2002,Shoot2002JCP,Chen2004,Oversteegen2004JPC}]
and the excellent monograph by Lekkerkerker and Tuinier \cite{Lek2011Book}). The
simulated phase boundaries of our HSC-HS mixture will be compared with the
theoretical corresponding predictions obtained with the one-fluid PL and
two-fluid MFP approaches. A simulation cell in a low-density thermodynamic state
is slowly compressed and equilibrated to obtain the dependence of bulk packing
fraction and bulk composition on the equilibrium bulk pressure (which for our
system with planar symmetry also corresponds to the normal component of the
pressure tensor, $P_n$). The bulk values of the phases are again obtained as
averages of the density and composition profiles in the central region of the
simulation cell.

\begin{figure}[htbp]
\centering
\includegraphics[angle=0,scale=0.45]{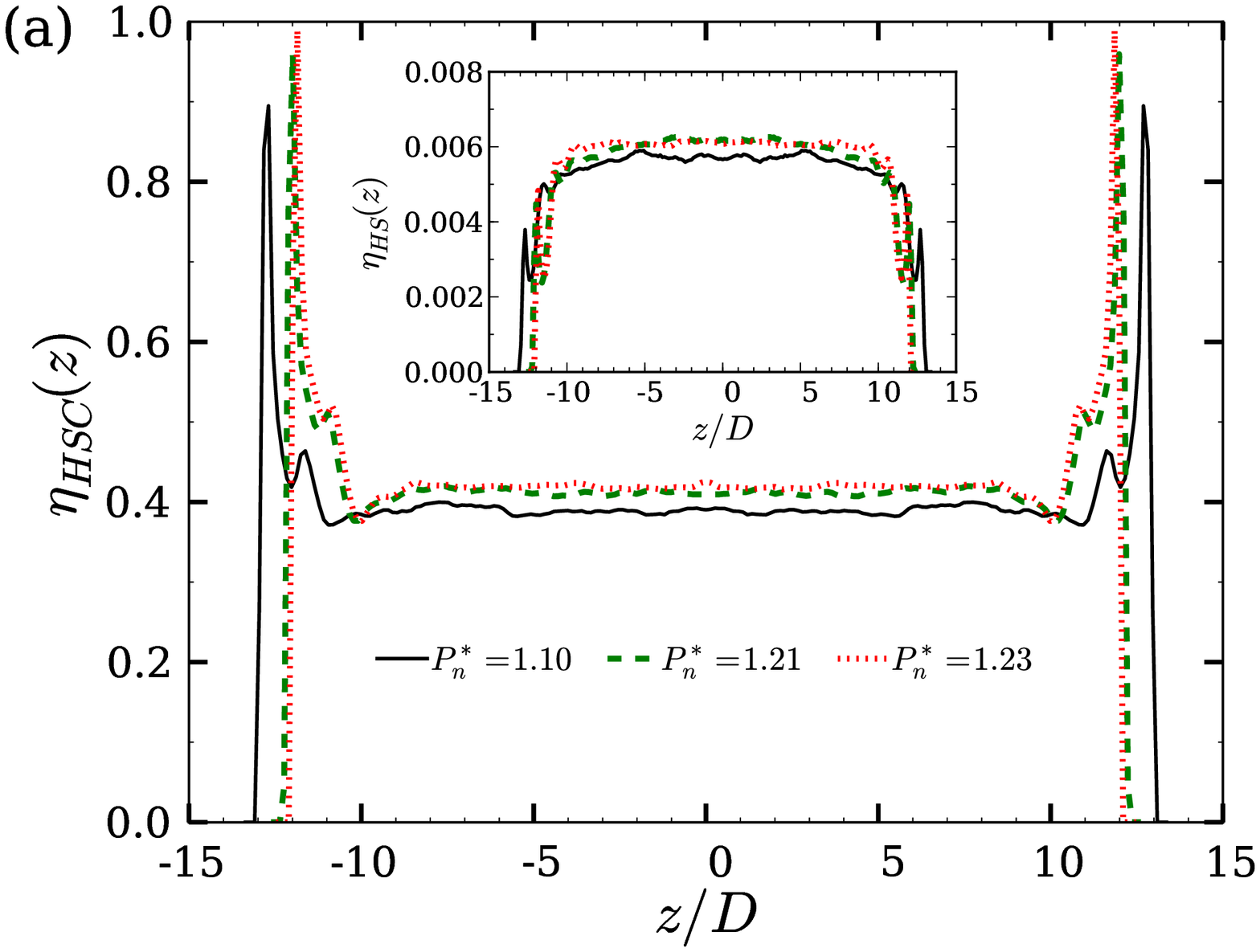}
\includegraphics[angle=0,scale=0.45]{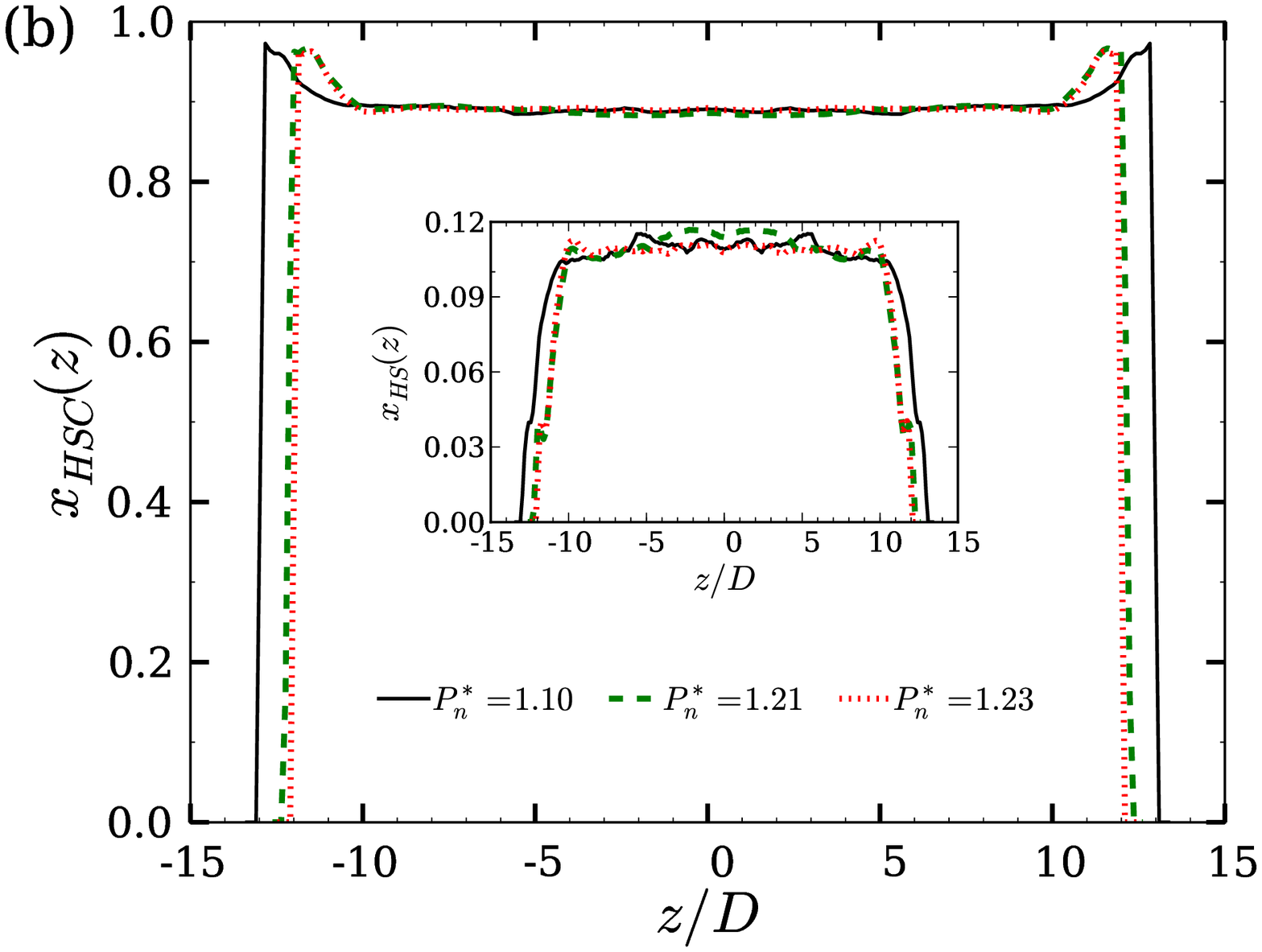}
\includegraphics[angle=0,scale=0.45]{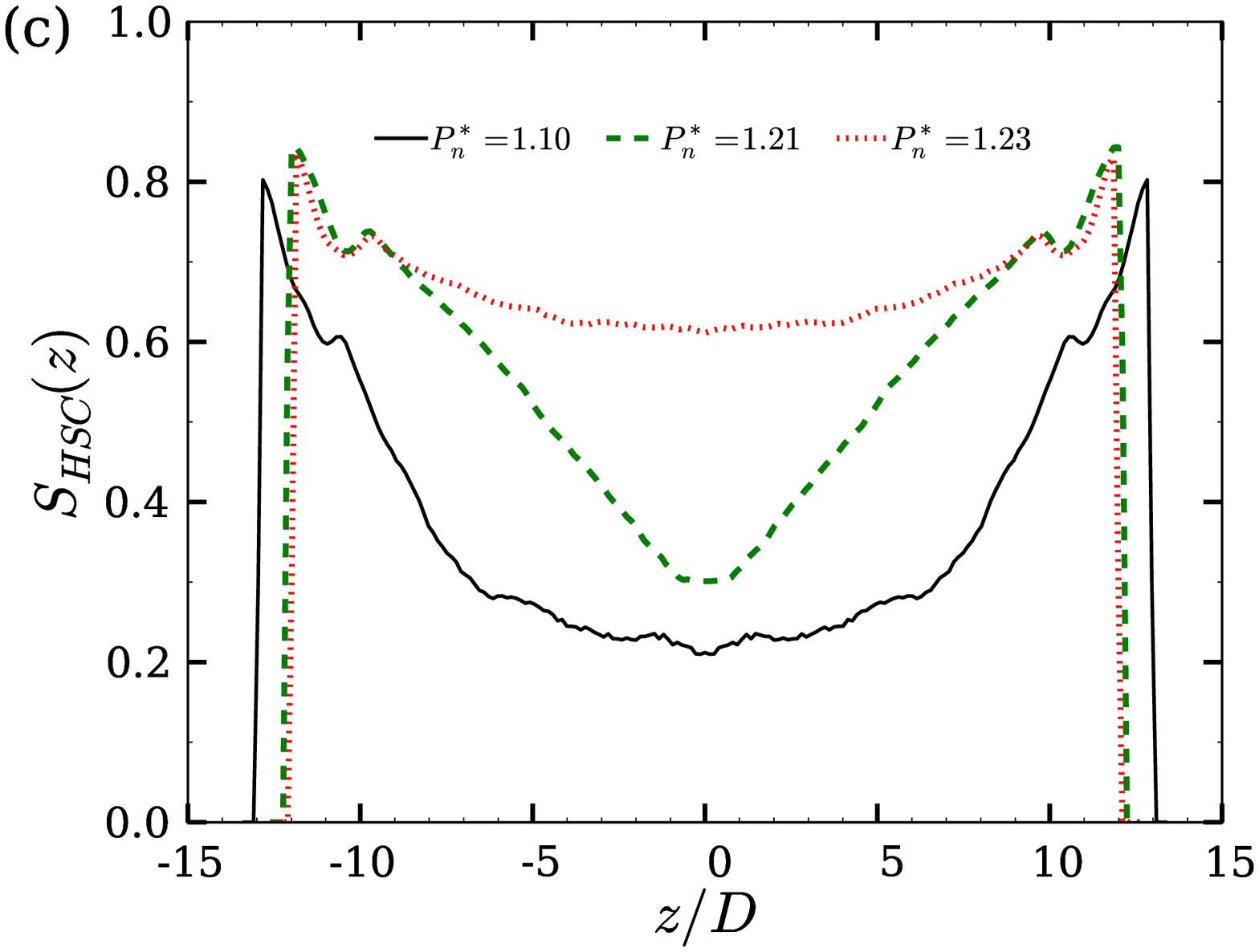}
\caption{(a) The packing fraction $\eta_{i}(z)$, (b) composition $x_{\rm i}(z)$, and  (c)
nematic order parameter $S_{2}(z)$ profiles
for mixtures of $N_{\rm HSC}=1482$ hard-spherocylinder (HSC) and $N_{\rm HS}=165$
  hard-sphere (HS) particles
for an overall composition of $x_{\rm HS,tot}=0.10$.
Typical bulk isotropic ($P_n^{*}=1.10$), pre-transitional ($P_n^{*}=1.21$),
and nematic ($P_n^{*}=1.23$) states are examined.
The packing fraction and composition profiles profiles of HS particles are shown in
the insets of (a) and (b).}
\label{chp5_fig_hschs_xhs=10_Profile}
\end{figure}

\begin{figure}[htbp]
\centering
\includegraphics[angle=0,scale=0.45]{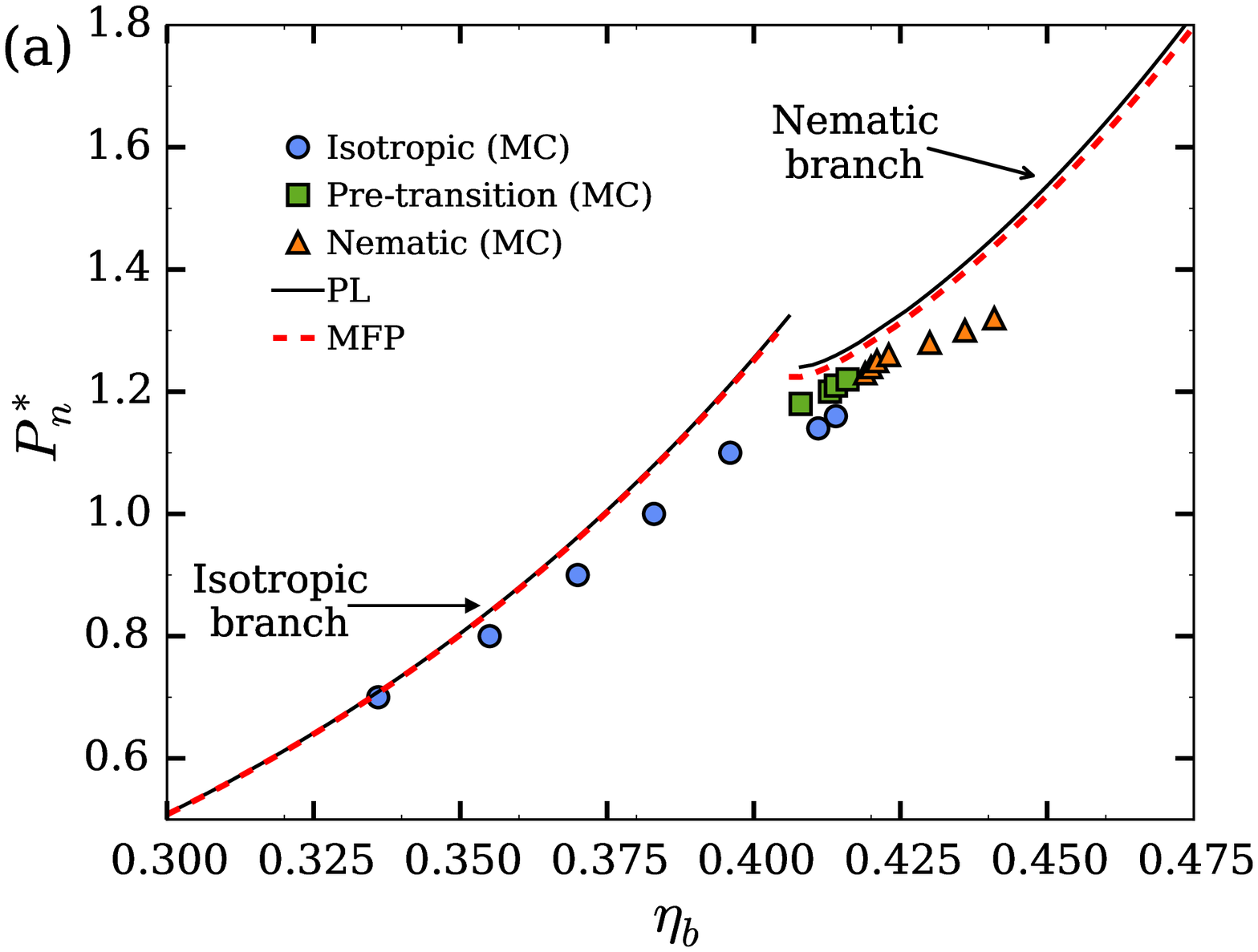}\\
\includegraphics[angle=0,scale=0.45]{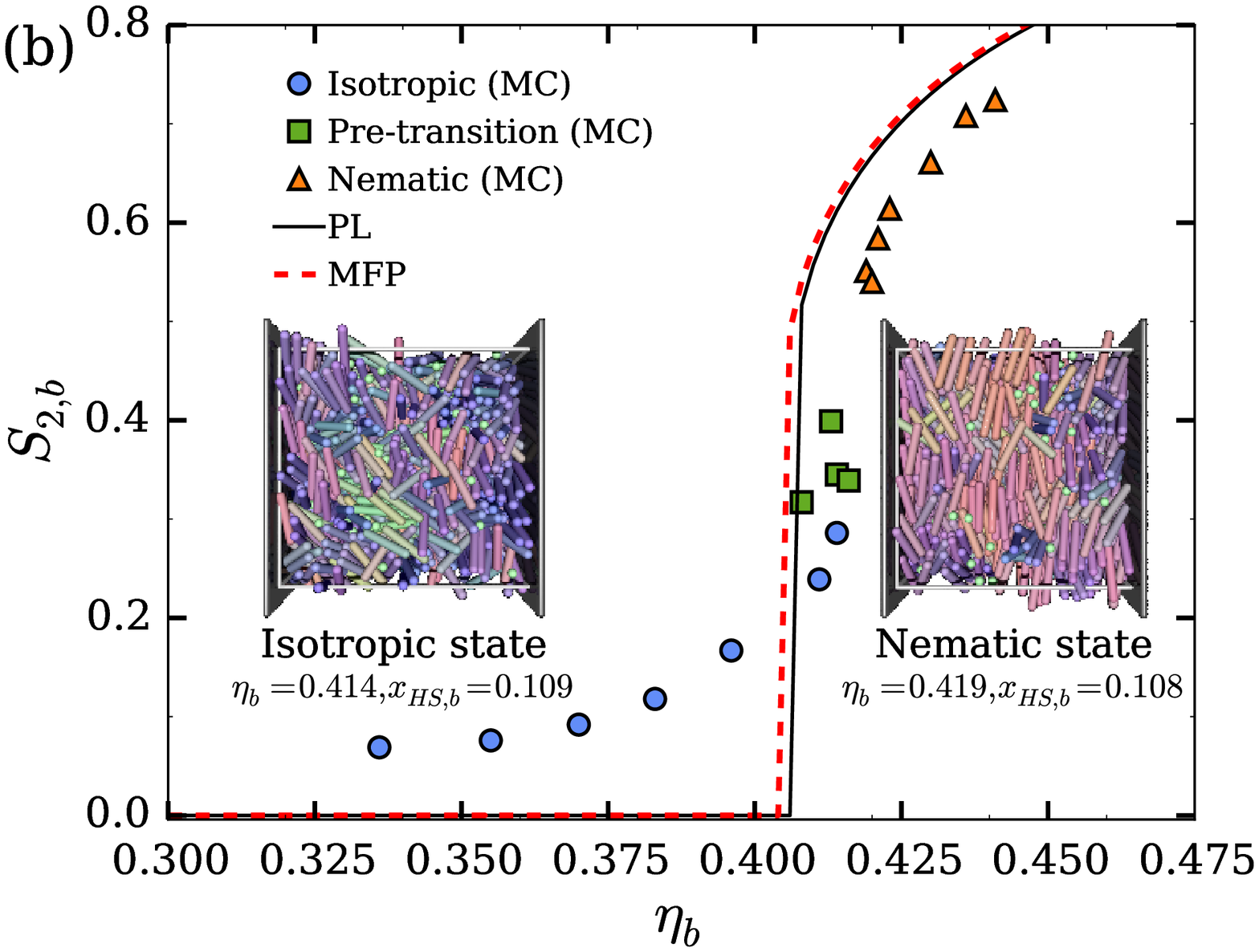}\\
\caption{(a) The pressure-density $P_n^{*}$-$\eta_{\rm b}$ phase diagram, and
(b) nematic orientational order parameter $S_{2,b}$ of mixtures of hard
spherocylinders (HSC) and hard spheres (HS). The results of $NP_nAT$-MC
simulations for mixtures of $N_{\rm HSC}=1482$ HSC and $N_{\rm HS}=165$ HS
particles for an overall HS composition of $x_{\rm HS,tot}=0.10$ contained
between well separated parallel hard walls are represented as: circles for the
isotropic states; triangles for the nematic states; and squares for the
pre-transitional states. The continuous and dashed curves represent the
predictions using the one-fluid PL and two-fluid MFP theories, respectively; the
low-density branch corresponds to isotropic states and the high-density branch
to nematic states. The snapshots in (b) correspond to the highest-density bulk
isotropic state ($\eta_{\rm b}=0.414, x_{\rm HS,b}=0.109$) and the
lowest-density bulk nematic state ($\eta_{\rm b}=0.419, x_{\rm HS,b}=0.108$).  }
\label{chp5_fig_MixBulk_Mix_EOS_S2_xhs10}
\end{figure}

Typical density, composition and nematic order parameter profiles for bulk isotropic,
intermediate isotropic-nematic, and bulk nematic states of mixtures with
overall HS composition $x_{\rm HS}=0.10$ are shown in Figure \ref{chp5_fig_hschs_xhs=10_Profile}, respectively.
As in the case of the pure-component HSC system, the density profiles of
the HSC-HS mixture reveal
significant order of the HSC close to the walls and a well-defined bulk region
characterised by the flat profiles in the central portion of the simulation cell.
This positional and orientational order is further confirmed by the large values
of composition and nematic order parameter  of the HSCs, and the low concentration of HSs
near the surface.
The characteristic flat profiles of the nematic order parameter in the bulk
isotropic at $P_n^*=1.10$ and nematic at $P_n^*=1.13$ states can be seen in
Fig. \ref{chp5_fig_hschs_xhs=10_Profile}(c); the V-shape nematic order profile of the
pre-transitional state  at $P^*_n=1.21$ is also clearly apparent.
The corresponding data for the mixture with 10\% HS is given in Table \ref{chp5_table_SimuData_HSCHS_0.1}.
The dependence of the bulk packing fraction $\eta_{\rm b}$
on the applied normal pressure $P_n^{*}$ for the HSC-HS system with an overall composition
of $x_{\rm HS,tot}=0.1$ is illustrated in Figure \ref{chp5_fig_MixBulk_Mix_EOS_S2_xhs10}(a).
The theoretical description of isotropic and nematic branches
of the equation of state obtained with the PL and MFP approach at the same overall composition
are also plotted in Figure \ref{chp5_fig_MixBulk_Mix_EOS_S2_xhs10}(a) for comparison.
There is only a negligible difference between the PL and MFP results in the isotropic state.
This is to be expected, since both the density and the ordering in this region is small and
as a consequence of the effective hard-sphere treatment of the excluded volume contributions with
a one- or two-fluid approximation should be similar for the isotropic state.
One should note that while within the theories (both PL and MFP) the density and
composition are input variables for a given system and the pressure is an output,
while for our $NP_nAT$-MC simulation the equilibrium pressure is specified and
the bulk density and composition are obtained as averages from the central region of the cell.
An observable difference between the one- and two-fluid theories can be seen
in the vicinity of the isotropic-nematic transition where
the branches of the equation of state obtained by simulation data experience
an abrupt change in slope indicating the transition to the orientationally ordered state.
From the results depicted in Figure \ref{chp5_fig_MixBulk_Mix_EOS_S2_xhs10},
one can infer that predictions with the two-fluid MFP approach is marginally superior to
that with the one-fluid PL at least for the nematic phase.
The isotropic-nematic transition point can be identified
from the abrupt change in nematic order parameter as is apparent from
Figure \ref{chp5_fig_MixBulk_Mix_EOS_S2_xhs10}(b).
The typical snapshots of configurations for the highest-density bulk isotropic state
($\eta_{\rm b}=0.414, x_{\rm HS,b}=0.109$)
and the lowest-density bulk nematic state ($\eta_{\rm b}=0.419, x_{\rm HS,b}=0.108$) are also included
in order to visualise differing degrees of orientational order.

\begin{table}[htbp]
\caption{Constant normal-pressure MC ($NP_nAT$-MC) simulation results for bulk
isotropic-nematic phase behaviour of mixtures of $N_{\rm HSC}=1482$ HSC
and $N_{\rm HS}=165$ HS particles for an overall HS composition of $x_{\rm HS,tot}=0.10$.
The reduced normal pressure $ P_n^{*}$ is set in the simulation and corresponding
bulk values of packing fraction $\eta_{\rm b}$, composition $x_{\rm HS, b}$,
nematic order parameter $S_{2,\rm b}$, and box length $L_z$.
are obtained as configurational averages. The isotropic phase is denoted by Iso,
the nematic by Nem, and the pre-transitional states by Pre.}
\centering
\begin{tabular*}{1.00\textwidth}{@{\extracolsep{\fill}} cccccc}
\toprule
$ P_n^{*}$  & $\eta_{\rm b}$  &$x_{\rm HS, b}$ & $S_{2,\rm b}$ & $ L_{z}/D$  & Phase \\
\colrule
  0.70 &  0.336 &  0.104 & 0.069 & 32.07 & iso\\
  0.80 &  0.355 &  0.105 & 0.076 & 30.34 & iso\\
  0.90 &  0.370 &  0.107 & 0.092 & 28.58 & iso\\
  1.00 &  0.383 &  0.109 & 0.118 & 27.56 & iso\\
  1.10 &  0.396 &  0.109 & 0.167 & 26.71 & iso\\
  1.14 &  0.411 &  0.109 & 0.239 & 26.15 & iso\\
  1.16 &  0.414 &  0.109 & 0.286 & 25.84 & iso\\
  1.18 &  0.408 &  0.115 & 0.317 & 25.65 & pre\\
  1.20 &  0.413 &  0.114 & 0.399 & 25.3  & pre\\
  1.21 &  0.414 &  0.114 & 0.345 & 25.35 & pre\\
  1.22 &  0.416 &  0.114 & 0.339 & 25.27 & pre\\
  1.23 &  0.419 &  0.108 & 0.551 & 25.18 & nem\\
  1.24 &  0.420 &  0.108 & 0.540 & 24.97 & nem\\
  1.25 &  0.421 &  0.112 & 0.584 & 24.66 & nem\\
  1.26 &  0.423 &  0.121 & 0.614 & 24.58 & nem\\
  1.28 &  0.430 &  0.111 & 0.661 & 24.35 & nem\\
  1.30 &  0.436 &  0.11  & 0.708 & 24.13 & nem\\
  1.32 &  0.441 &  0.109 & 0.724 & 23.89 & nem\\
\botrule
\label{chp5_table_SimuData_HSCHS_0.1}
\end{tabular*}
\end{table}

\begin{table}[htbp]
\caption{Constant normal-pressure MC ($NP_nAT$-MC) simulation results for bulk
isotropic-nematic phase behaviour of mixtures of $N_{\rm HSC}=1482$ HSC
and $N_{\rm HS}=371$ HS particles for an overall HS composition of $x_{\rm HS,tot}=0.20$.
The reduced normal pressure $ P_n^{*}$ is set in the simulation and corresponding
bulk values of packing fraction $\eta_{\rm b}$, composition $x_{\rm HS, b}$,
nematic order parameter $S_{2,\rm b}$, and box length $L_z$
are obtained as configurational averages. The isotropic phase is denoted by Iso,
the nematic by Nem, and the pre-transitional states by Pre.}
\centering
\begin{tabular*}{1.00\textwidth}{@{\extracolsep{\fill}} cccccc}
\toprule
$ P_n^{*}$ &  $\eta_{\rm b}$ & $x_{\rm HS, b}$ & $S_{2,\rm b}$ & $ L_{z}/D$ & Phase \\
\colrule
   0.80 & 0.348 & 0.206 & 0.131 & 31.05 &  Iso \\
   0.90 & 0.361 & 0.211 & 0.151 & 29.75 &  Iso \\
   1.00 & 0.379 & 0.212 & 0.172 & 28.68 &  Iso \\
   1.10 & 0.391 & 0.213 & 0.163 & 27.85 &  Iso \\
   1.20 & 0.403 & 0.222 & 0.192 & 26.45 &  Iso \\
   1.22 & 0.403 & 0.224 & 0.193 & 26.21 &  Iso \\
   1.24 & 0.404 & 0.220 & 0.260 & 25.95 &  Iso \\
   1.26 & 0.408 & 0.226 & 0.301 & 25.62 &  Pre \\
   1.28 & 0.414 & 0.225 & 0.404 & 25.53 &  Pre \\
   1.30 & 0.415 & 0.230 & 0.415 & 25.35 &  Pre \\
   1.31 & 0.417 & 0.233 & 0.419 & 25.26 &  Pre \\
   1.32 & 0.430 & 0.219 & 0.619 & 25.02 &  Nem \\
   1.34 & 0.430 & 0.223 & 0.612 & 24.71 &  Nem \\
   1.36 & 0.431 & 0.226 & 0.628 & 24.50 &  Nem \\
   1.38 & 0.432 & 0.226 & 0.619 & 24.64 &  Nem \\
   1.39 & 0.436 & 0.226 & 0.650 & 24.53 &  Nem \\
\botrule
\label{chp5_table_SimuData_HSCHS_0.2}
\end{tabular*}
\end{table}

\begin{figure}[htbp]
\centering
\includegraphics[angle=0,scale=0.45]{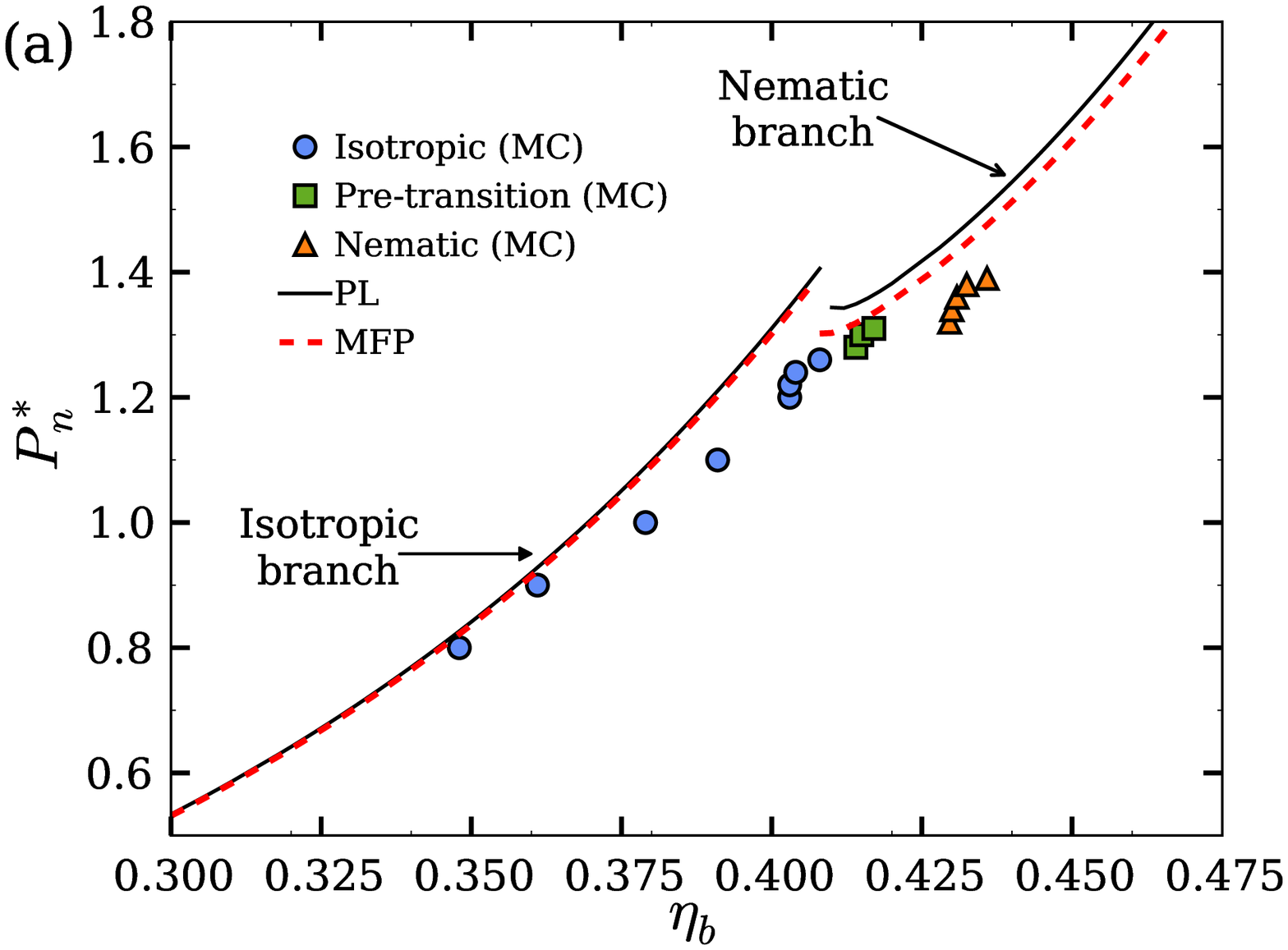}\\
\includegraphics[angle=0,scale=0.45]{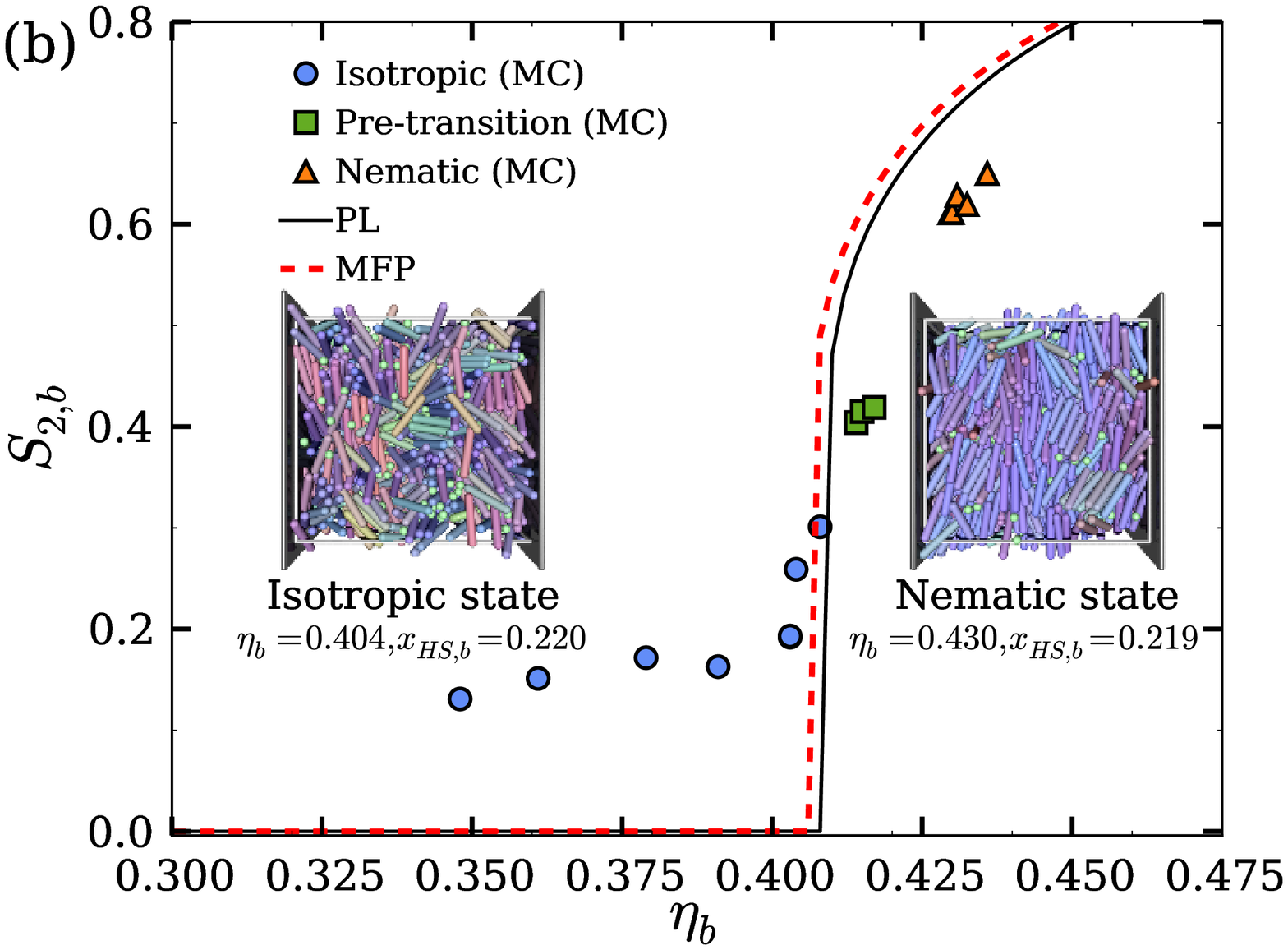}\\
\caption{(a) The pressure-density $P_n^{*}$-$\eta_{\rm b}$ phase diagram and (b) nematic 
orientational order parameter $S_{2,b}$ of mixtures of hard spherocylinders (HSC) and hard
spheres (HS) particles. The results of $NP_nAT$-MC simulations for mixtures of $N_{\rm HSC}=1482$ HSC and $N_{\rm HS}=371$ HS particles for an overall HS composition of
$x_{\rm HS,tot}=0.20$ contained between well separated parallel hard walls are represented as: circles for the isotropic states; triangles for the nematic states; and
squares for the pre-transitional states. The continuous and dashed curves represent the predictions using the one-fluid PL and two-fluid MFP theories, respectively;
the low-density branch corresponds to isotropic states and the high-density branch to nematic states. The snapshots in (b) correspond to the highest-density bulk
isotropic state ($\eta_{\rm b}=0.404, x_{\rm HS,b}=0.220$) and the lowest-density bulk nematic state ($\eta_{\rm b}=0.430, x_{\rm HS,b}=0.219$). }
\label{chp5_fig_MixBulk_Mix_EOS_S2_xhs20}
\end{figure}

The corresponding results for the HSC-HS system with the highest overall HS composition $x_{\rm HS,tot}=0.20$,
are shown in Figure \ref{chp5_fig_MixBulk_Mix_EOS_S2_xhs20} and in Table \ref{chp5_table_SimuData_HSCHS_0.2}.
The isotropic-nematic
transition lies somewhere between highest-density bulk isotropic state at $P_n^{*}=1.24$ and
the lowest-density bulk nematic state at $P_n^{*}=1.32$ where a clear change
in the density and nematic order parameter is exhibited:
the values of the packing fraction and nematic order parameter for these two states are
$\eta_{\rm b}=0.404$ and $S_{2,b}=0.260$ for the bulk isotropic phase and
$\eta_{\rm b}=0.430$ and $S_{2,b}=0.619$ for the bulk nematic phase.
For completeness the less extensive data for overall hard-sphere compositions 
of 5\% and 15\% 
are given in Tables \ref{Table_x0.05} and \ref{Table_x0.15}.

\begin{figure}[htbp]
\centering
\includegraphics[angle=0,scale=0.45]{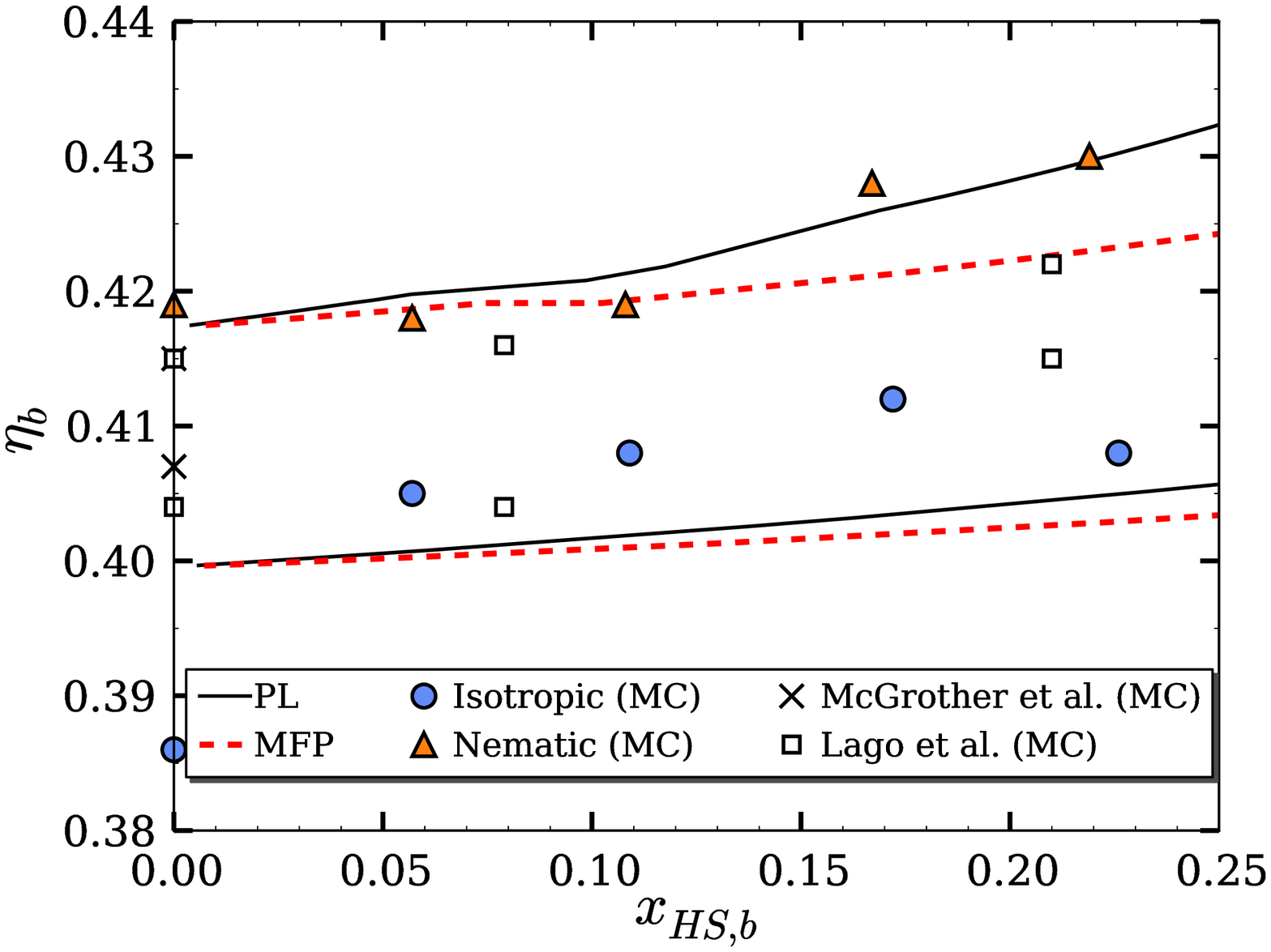}\\
\caption{\label{chp5_fig_MixBulk_Phase_MixEtaXhs}The isotropic-nematic phase
diagram of mixtures of hard spherocylinders (HSC) and hard spheres (HS) particles in
$\eta_{\rm b} (\eta)$-$x_{\rm HS,b}$ plane.   The results of $NP_nAT$-MC
simulations for mixtures of $N_{\rm HSC}=1482$ HSC and $N_{\rm HS}=165$ HS
particles for an overall HS composition of $x_{\rm HS,tot}=0.1$ are represented
as filled circles for the isotropic branch and filled triangles for the nematic
branch. The $NP_nAT$-MC data are compared with predictions obtained with the
one-fluid PL (dashed curves) and two-fluid MFP (continuous curves) theories, and
with previously reported $NPT$-MC simulation data for fully periodic systems:
open squares \cite{Lago2004JMR} and crosses \cite{McGrother1996}.}
\end{figure}

The isotropic-nematic phase boundaries estimated for the HSC-HS mixtures
for bulk phase compositions ranging from the pure HSC system ($x_{HS,b}=0$)
to $x_{HS,b}\sim 0.3$ is shown in Figure \ref{chp5_fig_MixBulk_Phase_MixEtaXhs}.
Here, we also compare the predictions of PL and MFP theories with our $NP_nAT$-MC simulations
as well as with the previously reported $NPT$-MC data for systems
in full three-dimensional periodic boundary conditions \cite{Lago2004JMR}.
The one-fluid and two-fluid approaches are both seen to describe the coexisting packing fractions as monotonically
increasing functions of the bulk composition for the isotropic and nematic phases.
For the phase boundary of the nematic states found at the higher packing fractions,
the difference between the description obtained with the PL and MFP approaches
becomes more marked with increasing composition of the spherical particles.
Our simulated values for the nematic and isotropic phase boundaries are consistent with those obtained with a fully periodic system;
the simulation data are in reasonably good agreement with the PL and MFP theoretical descriptions.
There is an overestimate of the first-order character
of the isotropic-nematic transition with scaled Onsager theories of this type
 based on a underlying description at the level of the second virial coefficient \cite{McGrother1996}.
The approximations inherent in mapping HSC-HS mixture on to an equivalent HS mixture
in order to simplify the computation of higher-order virial contribution
may also lead to an exaggeration of the first-order nature of the transition.
On increasing the overall composition of the HS particles,the density gap between
the isotropic and nematic coexisting states is seen to become larger as found with our simulations.
By contrast,the density gap between the phase boundaries obtained from
conventional $NPT$-MC simulations of the fully periodic system \cite{Lago2004JMR} appears to shrink slightly.
In fully periodic $NPT$-MC simulations of this type the system remains essentially homogeneous
so that the composition of the state remains fixed. As the pressure is increased the system will
undergo a transition from an isotropic to nematic phase but the states will be constrained to have the same composition.
As a consequence of lack of fractionation of the species between the two phases with the
$NPT$-MC simulations it is difficult to differentiate metastable states within
the binodal region from those corresponding to the coexistence boundaries.

The $NP_nAT$-MC simulation data for the HSC-HS mixture are also summarized reported in Table \ref{chp5_table_Mixbulk_IsoNem}
where the slight composition asymmetry between coexisting isotropic and nematic phases is clearly apparent.
For the HSC-HS system, the addition of spherical particles is found to destabilize.
The formation of a bulk nematic state predominantly composed of HSC rods will cause
a reduction in the concentration of the HS particles in the same region,
and as a consequence the orientational ordered state which is of higher density
but lower bulk HS composition coexists with an isotropic state of lower-density and higher bulk HS composition.

\begin{figure}[htbp]
\centering
\includegraphics[angle=0,scale=0.45]{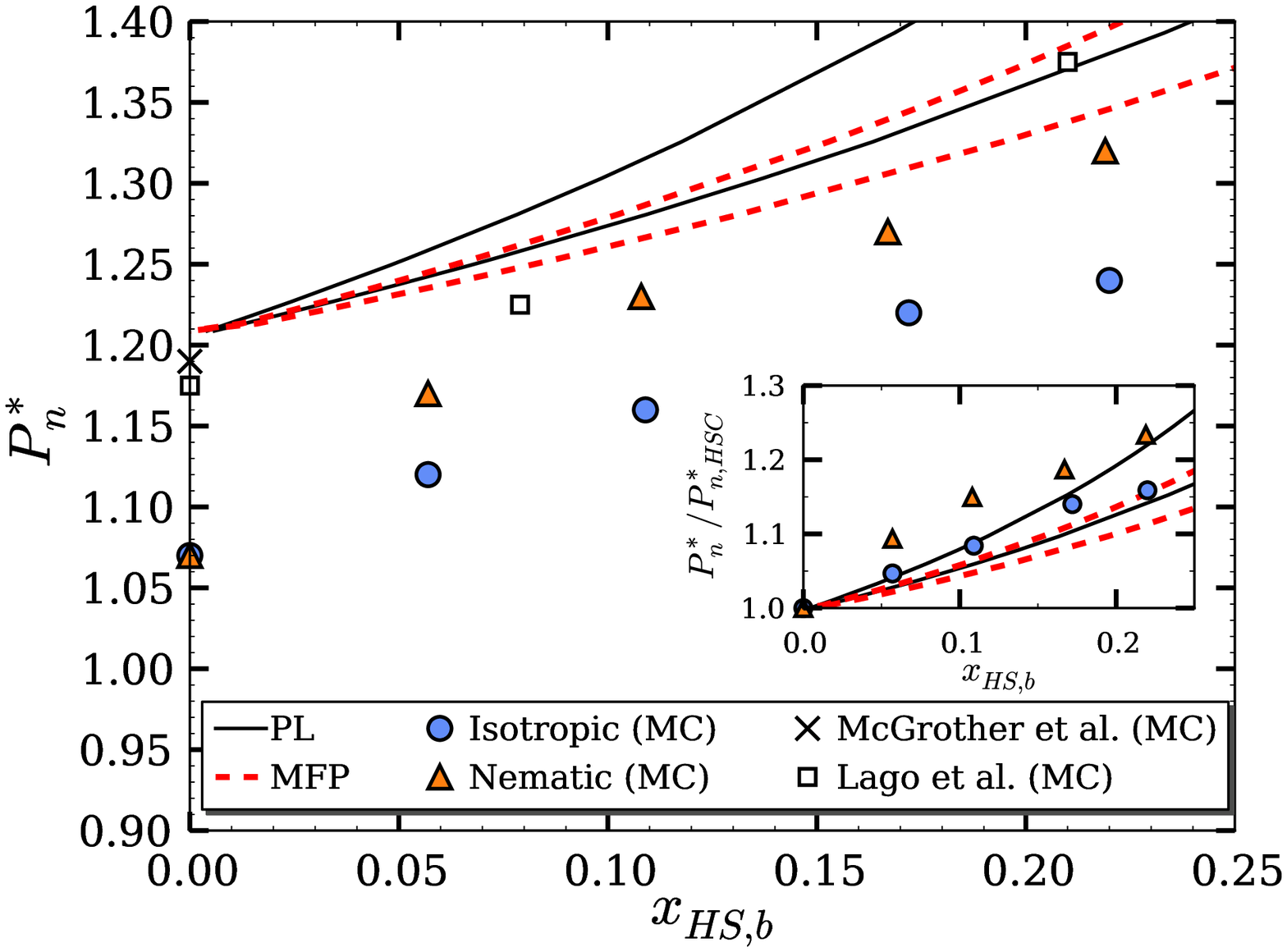}\\
\caption{\label{chp5_fig_MixBulk_Phase_MixPrsXhs} The isotropic-nematic phase
diagram of mixtures of hard spherocylinders (HSC) and hard spheres (HS) particles in
$P^{*}_{n} $-$x_{\rm HS,b}$ plane. The results of $NP_nAT$-MC
simulations for mixtures of $N_{\rm HSC}=1482$ HSC and $N_{\rm HS}=165$ HS
particles for an overall HS composition of $x_{\rm HS,tot}=0.1$ are represented
as filled circles for the isotropic branch and filled triangles for the nematic
branch. The $NP_nAT$-MC data are compared with predictions obtained with the
one-fluid PL (dashed curves) and two-fluid MFP (continuous curves) theories, and
with previously reported $NPT$-MC simulation data for fully periodic systems:
open squares \cite{Lago2004JMR} and crosses \cite{McGrother1996}. The simulation
results obtained from $NP_nAT$ simulations have been rescaled with the
pure component pressure in the inset; the same rescaling
procedure applied to the results obtained using the PL and MFP
theories.}
\end{figure}

Finally, a phase diagram in the pressure-composition ($P_n^{*}$-$x_{\rm HS,b}$) plane is shown in Figure \ref{chp5_fig_MixBulk_Phase_MixPrsXhs}. A very narrow region
of isotropic-nematic coexistence is obtained with our $NP_nAT$-MC simulation approach. The coexistence region is seen to be at lower pressures that that predicted
with the PL and MFP theories or obtained by fully periodic $NPT$-MC simulation \cite{Lago2004JMR}. It should be noted that the results obtained at higher compositions
with the fully periodic simulations is in good agreement with the nematic branch predicted with MFP theory. The quality of both PL and MFP is affected by a shift in
pressure for the pure HSC system. The predictions with the two-fluid MFP theory are seen to be much better than one-fluid PL theory, particularly for systems of
higher composition.

\begin{table}[htbp]
\caption{Constant normal-pressure MC ($NP_nAT$-MC) simulation results for bulk
isotropic-nematic phase behaviour of mixtures of $N_{\rm HSC}=1482$ HSC
and $N_{\rm HS}=78$ HS particles for an overall HS composition of $x_{\rm HS,tot}=0.05$.
The reduced normal pressure $ P_n^{*}$ is set in the simulation and corresponding
bulk values of packing fraction $\eta_{\rm b}$, composition $x_{\rm HS, b}$,
nematic order parameter $S_{2,\rm b}$, and box length $L_z$
are obtained as configurational averages. The isotropic phase is denoted by Iso,
the nematic by Nem, and the pre-transitional states by Pre.}
\centering
\begin{tabular*}{1.00\textwidth}{@{\extracolsep{\fill}} cccccc}
\toprule
$ P_n^{*}$ &  $\eta_{\rm b}$ &  $x_{\rm HS, b}$ &$S_{2,\rm b}$ & $ L_{z}/D$  & Phase \\
\colrule
   0.700 & 0.337 &  0.052 &  0.072 &  31.380 & Iso \\
   0.800 & 0.355 &  0.054 &  0.080 &  29.610 & Iso \\
   0.900 & 0.372 &  0.054 &  0.100 &  28.210 & Iso \\
   1.000 & 0.382 &  0.056 &  0.123 &  27.190 & Iso \\
   1.100 & 0.401 &  0.057 &  0.245 &  25.860 & Iso \\
   1.120 & 0.405 &  0.057 &  0.298 &  25.630 & Iso \\
   1.140 & 0.409 &  0.059 &  0.345 &  25.330 & Pre \\
   1.150 & 0.416 &  0.058 &  0.392 &  25.330 & Pre \\
   1.160 & 0.412 &  0.061 &  0.392 &  25.270 & Pre \\
   1.170 & 0.418 &  0.057 &  0.574 &  24.960 & Nem \\
   1.180 & 0.424 &  0.056 &  0.657 &  24.540 & Nem \\
   1.200 & 0.429 &  0.062 &  0.641 &  24.520 & Nem \\
\botrule
\label{Table_x0.05}
\end{tabular*}
\end{table}

\begin{table}[htbp]
\caption{Constant normal-pressure MC ($NP_nAT$-MC) simulation results for bulk
isotropic-nematic phase behaviour of mixtures of $N_{\rm HSC}=1482$ HSC
and $N_{\rm HS}=262$ HS particles for an overall HS composition of $x_{\rm HS,tot}=0.15$.
The reduced normal pressure $ P_n^{*}$ is set in the simulation and corresponding
bulk values of packing fraction $\eta_{\rm b}$, composition $x_{\rm HS, b}$,
nematic order parameter $S_{2,\rm b}$, and box length $L_z$
are obtained as configurational averages. The isotropic phase is denoted by Iso,
the nematic by Nem, and the pre-transitional states by Pre.}
\centering
\begin{tabular*}{1.00\textwidth}{@{\extracolsep{\fill}} cccccc}
\toprule
$ P_n^{*}$ &  $\eta_{\rm b}$  & $x_{\rm HS, b}$ & $S_{2,\rm b}$ & $ L_{z}/D$  & Phase \\
\colrule
  1.000 &  0.381 &  0.163&  0.151  &  27.948 &  Iso \\
  1.100 &  0.392 &  0.166&  0.180  &  26.819 &  Iso \\
  1.200 &  0.412 &  0.169&  0.260  &  25.524 &  Iso \\
  1.210 &  0.406 &  0.174&  0.241  &  25.645 &  Iso \\
  1.220 &  0.412 &  0.172&  0.306  &  25.757 &  Iso \\
  1.250 &  0.421 &  0.171&  0.360  &  25.290 &  Pre \\
  1.260 &  0.420 &  0.171&  0.441  &  25.104 &  Pre \\
  1.270 &  0.428 &  0.167&  0.612  &  24.805 &  Nem \\
  1.280 &  0.424 &  0.170&  0.560  &  24.990 &  Nem \\
\botrule
\label{Table_x0.15}
\end{tabular*}
\end{table}

\begin{table}[htbp]
\caption{The isotropic-nematic transition estimated from $NP_nAT$-MC simulations of
mixtures of $N_{\rm HSC}=1482$ hard spherocylinder (HSC) and $N_{\rm HS}$ hard-sphere (HS) particles
for varying overall HS compositions of $x_{\rm HS,tot}$ contained between well separated parallel hard walls.
The normal pressure $P_n^{*}=P_{n}D^3/(k_{\rm B} T)$ is set during the simulation,
bulk packing fractions $\eta_{\rm b}$ and bulk compositions $x_{\rm HS,b}$ of coexisting
isotropic (iso) and nematic (nem) states are obtained as averages from the central region of the cell.
The bulk nematic order parameter $S_{2,\rm b}$ of the lowest-density nematic bulk phase is also shown.}
\centering
\begin{tabular*}{1.00\textwidth}{@{\extracolsep{\fill}} cccccccc}
\toprule
$x_{\rm HS,tot}$  & $ P_{n,\rm iso}^{*}$  & $\eta_{\rm b,iso}$ & $x_{\rm HS,b,iso}$
                  & $ P_{n,\rm nem}^{*}$  & $\eta_{\rm b,nem}$ & $x_{\rm HS,b,nem}$ & $S_{2,\rm b}$ \\
\colrule
0    & 1.00  & 0.386  & 0      & 1.14 & 0.419  & 0       & 0.553 \\
0.05 & 1.12  & 0.405  & 0.057  & 1.17 & 0.418  & 0.057   & 0.574 \\
0.10 & 1.16  & 0.414  & 0.109  & 1.23 & 0.419  & 0.108   & 0.551 \\
0.15 & 1.22  & 0.412  & 0.172  & 1.27 & 0.428  & 0.167   & 0.612 \\
0.20 & 1.26  & 0.408  & 0.226  & 1.32 & 0.430  & 0.219   & 0.619 \\
\botrule
\label{chp5_table_Mixbulk_IsoNem}
\end{tabular*}
\end{table}

\section{Conclusions} \label{chp5_Sec_Conclusion}

In this paper we have studied phase behaviour of mixtures of purely repulsive rod-like
(HSC) and spherical (HS) particles. Using the new $NP_nAT$ Monte Carlo simulations we have
constructed the isotropic-nematic phase diagrams of the HSC-HS mixture in the
pressure-density, pressure-concentration and density-concentration projections. The
comparison of our results with previously reported fully periodic $NPT$-MC data
\cite{Lago2004JMR} reveals good agreement at low HS concentrations but some discrepancy
occurs for higher HS concentrations especially in the nematic phase.  In conventional
$NPT$-MC simulations the system is essentially homogeneous such that the overall
composition of the system remains fixed. This means that unless one has a very large
system the coexisting isotropic and nematic states would have identical compositions
\cite{khasm13,thajcp14}.  By contrast, using our $NP_nAT$-MC algorithm we have observed
compositional asymmetry between the coexisting phases, with a slight increase of the HSC
concentration in the nematic phase due to packing effects. This conclusion is also
supported by the fact that our $NP_nAT$ results for pure HSC particles are completely consistent
with the available data obtained from $NPT$-MC simulations using full three-dimensional
boundary conditions \cite{McGrother1996}.

The advantage of our method is that the fluid phase separation can be studied within a single simulation cell which circumvents the problem of inserting anisotropic
particles inherent in other techniques such as Gibbs ensemble MC (GEMC). Additionally, particle exchanges are allowed between the surface and bulk regions so that one
is able to treat the two bulk states coexisting at different bulk densities (packing fractions) as well as different bulk compositions. The effect of the pair of the
auxiliary hard walls put at the end of the box is not appreciable in the bulk region for sufficiently prolonged systems in the direction perpendicular to the walls
and considerably helps to stabilise the phase separated states with a low interfacial tension which as typically exhibited by hard-body systems.

We further compared the new simulation data with two theoretical predictions that go
beyond the Onsager second-virial theory and extend the well known PL theory to mixtures in
two different ways: using the one-fluid approximation whereby one maps the mixture onto an
effective HS mixture, and by treating both species separately as  an
effective HS mixture which we refer to as many-fluid approximation. The comparison reveals
that both the one- and many-fluid approaches provide a reasonably accurate quantitative
description of the mixture including  the isotropic-nematic phase
boundary and degree of orientational order of the HSC-HS mixtures. The many-fluid
prediction of the coexisting pressure is arguably found to be slightly better to that obtained
with the one-fluid method. However, systems with larger aspect ratios should be considered
to make a better assessment of the performance of the theories.

Our work can be directly extended in a number of ways. For instance, we have restricted
our attention to systems of HSC rods with an aspect ratio of $L/D=5$ and HS particles of
diameter $\sigma=D$. Considering larger aspect ratios would be a more stringent test of
the theoretical predictions, as one would expect a more significant improvement of
the many-fluid treatment compared to the one-fluid approach. The compositional
asymmetry in the coexisting isotropic and nematic phases is expected to be more
appreciable. The method for locating the phase boundaries can be improved using
thermodynamic integration. Furthermore, our $NP_nAT$-MC approach can be directly applied
to describe the surface phenomena due to both fluid-fluid and fluid-wall
interfaces. These interfaces plays an important role in liquid crystalline systems
\cite{Mao1997MP,Dijkstra2001PRE} due of rich surface-induced effects (e.g., nematization
and smectization), characteristic in systems comprising  anisotropic particles.

\begin{acknowledgments}

LW thanks Department for Business Innovation and Skills, UK and China Scholarship Council for funding a PhD studentship. MC simulation in this work is performed using
the High Performance Computing service provided by Imperial College London. AM acknowledges a support from the Czech Science Foundation, Grant No. 13-02938S. Funding
to the Molecular Systems Engineering Group from the Engineering and Physical Sciences Research Council (EPSRC) 
of the U.K. (grants GR/T17595, GR/N35991, EP/E016340, and EP/J014958), the Joint Research Equipment Initiative (JREI) (GR/M94426), and the Royal Society-Wolfson Foundation refurbishment scheme is also gratefully
acknowledged.

\end{acknowledgments}

%

%\bibliographystyle{apsrev4-1}
%\bibliography{bibli.bib}

\end{document}